\newdimen\bibsep
\documentclass[3p,nonatbib]{elsarticle}

\usepackage{amssymb}
\usepackage{amsmath}
\usepackage{booktabs}
\usepackage{algpseudocode}
\usepackage{graphicx}
\graphicspath{ {./} {./figs/} } 
\usepackage{subfig}
\usepackage[inkscapelatex=false]{svg}
\usepackage{nomencl}
\usepackage{tikz}
\usepackage[normalem]{ulem}
\usepackage{xcolor} 
\usepackage{comment}

\makeatletter
\@ifundefined{c@author}{}{\let\c@author\relax}
\@ifundefined{bibfont}{}{}
\@ifundefined{Citeauthor}{}{}
\@ifundefined{citename}{}{}
\makeatother

\usepackage[
  backend=biber,
  style=numeric-comp, 
  sorting=none,
  sortcites=true,
  maxnames=99,
  giveninits=true
]{biblatex}

\makenomenclature

\usepackage{etoolbox}
\renewcommand\nomgroup[1]{%
  \item[\bfseries
  \ifstrequal{#1}{A}{Acronyms}{%
  \ifstrequal{#1}{P}{Physical Terms}{%
  \ifstrequal{#1}{G}{Greek Symbols}{%
  \ifstrequal{#1}{S}{Subscripts}{}}}}%
]}

\nomenclature[P]{$\dot Q$}{heat transfer rate (kW)}
\nomenclature[P]{$\dot W$}{power (kW)}
\nomenclature[P]{$x$}{chemical energy stored in battery (kWh)}
\nomenclature[P]{$\overline x$}{battery chemical energy capacity (kWh)}
\nomenclature[P]{$u$}{battery chemical charging power (kW)}
\nomenclature[P]{$b$}{battery electric charging power (kW)}
\nomenclature[P]{$\overline b$}{battery electric power capacity (kW)}
\nomenclature[P]{$s$}{solar power (kW)}
\nomenclature[P]{$d$}{heat pump power (kW)}

\nomenclature[A]{AC}{alternating current}
\nomenclature[A]{AHRI}{Air-Conditioning, Heating, and Refrigeration Institute}
\nomenclature[A]{DC}{direct current}
\nomenclature[A]{PV}{photovoltaic}
\nomenclature[A]{COP}{coefficient of performance}

\nomenclature[S]{th}{thermal}
\nomenclature[S]{e}{electric}
\nomenclature[S]{in,fan}{indoor unit fan}
\nomenclature[S]{comp}{compressor}
\nomenclature[S]{out,fan}{outdoor unit fan}

\nomenclature[P]{$\eta$}{battery charging and discharging efficiency (-)}
\nomenclature[P]{$a$}{discrete-time dynamics parameter}
\nomenclature[P]{$\tau$}{battery self-dissipation time constant}

\journal{Elsevier}

\begin{document}

\begin{frontmatter}

\title{Laboratory and field testing of a residential heat pump retrofit for a DC solar nanogrid}

\author[label1]{Aaron H.P. Farha}
\author[label2]{Jonathan P. Ore}
\author[label1,label4]{Elias N. Pergantis}
\author[label1]{Davide Ziviani}
\author[label1]{Eckhard A. Groll}
\author[label1,label3,*]{Kevin J. Kircher}

\affiliation[label1]{organization={Ray W. Herrick Laboratories, School of Mechanical Engineering, Purdue University},
            city={West Lafayette},
            state={IN},
            country={USA}}

\affiliation[label2]{organization={Emerson Automation Solutions},
            city={Pittsburgh},
            state={PA},
            country={USA}}

\affiliation[label3]{organization={Elmore Family School of Electrical and Computer Engineering},
            city={West Lafayette},
            state={IN},
            country={USA}}
            
\affiliation[label4]{organization={Trane Technologies, Residential R\&D Group},
            city={Tyler},
            state={TX},
            country={USA}}

\affiliation[*]{organization={Corresponding author}}

\begin{abstract}
Residential buildings are increasingly integrating large devices that run natively on direct current (DC), such as solar photovoltaics, electric vehicles, stationary batteries, and DC motors that drive heat pumps and other major appliances. Today, these natively-DC devices typically connect within buildings through alternating current (AC) distribution systems, entailing significant energy losses due to conversions between AC and DC. This paper investigates the alternative of connecting DC devices through DC distribution. Specifically, this paper shows through laboratory and field experiments that an off-the-shelf residential heat pump designed for conventional AC systems can be powered directly on DC with few hardware modifications and little change in performance. Supporting simulations of a DC nanogrid including {historical heat pump and rest-of-house load measurements,} a solar photovoltaic array, and a stationary battery suggest that connecting these devices through DC distribution could decrease annual electricity bills by 12.5\% with an after-market AC-to-DC heat pump retrofit and by 16.7\% with a heat pump designed to run on DC. The associated savings in gross nanogrid energy are 8\% and 9.2\%, respectively.
\end{abstract}

\begin{highlights}
\item Off-the-shelf residential air-source heat pump retrofit to run on DC

\item Experimental evaluation in psychrometric chambers and an occupied 208 m$^2$ test house

\item Laboratory and field experiments show similar heat pump performance on AC and DC

\item DC solar nanogrid model calibrated to empirical data from heat pump testing

\item {Model shows 12.5 to 16.7\% annual net electricity bill savings in DC vs. AC configuration}

\end{highlights}


\end{frontmatter}

\section{Introduction}
\label{intro}

Residential buildings use over one-third of United States electricity \cite{eia-2025-energy-use}. Many large residential loads, such as motors that drive compressors, pumps, and fans, run natively on DC power \cite{assaf_power_2024, garbesi_catalog_2010, stippich_ac_2017}. Residential buildings are increasingly incorporating other large, natively-DC devices, such as solar photovoltaics (PV), electric vehicle chargers, and stationary batteries. Today, these natively-DC devices connect within buildings through AC distribution systems, requiring conversions between AC and DC at efficiencies that typically range from 85 to 95\%  \cite{vossos_energy_2014}. Connecting natively-DC devices through DC distribution systems could reduce energy conversion losses, the installation and maintenance costs associated with inverters and rectifiers, and the use of materials such as copper and aluminum \cite{fotopoulou2021state, rodriguez2015overview}.

While several modeling studies suggest that connecting DC devices within residential buildings through DC distribution could have significant benefits \cite{vossos_energy_2014, chen_networked_2021}, {these studies often overlook key practical challenges associated with real-world implementation.} Thus, experimental work is needed to convince decision-makers in business and government that DC technologies are worth bringing to market. Toward that end, Charalambous et al. investigated a hybrid AC-DC nanogrid connecting PV, a battery, and a heat pump in a small building meant to exhibit renewable energy technologies \cite{CHARALAMBOUS-2023}. Charalambous et al. showed that reducing the number of AC-DC conversions raised PV self-consumption above 85\% {in a site study in Cyprus}. Ore et al. switched a 14 kW (4 ton) heat pump to a 350  V DC source and ran steady-state laboratory cooling tests at maximum compressor speed following Air-Conditioning, Heating, and Refrigeration Institute (AHRI) Standard 210/240 \cite{ore_evaluation_2020}. Ore et al. estimated that switching the heat pump to DC could reduce input energy by 6 to 8\%. Beyond heating and cooling, experimentalists have studied the switch to DC-driven light-emitting diode lighting systems. For example, Sriram et al. \cite{sriram_development_2024} and Chacko et al. \cite{chacko_dc_2022} showed that powering LEDs directly on DC could reduce input energy by up to 22\%.

Complementing past experimental studies, this paper investigates the impacts of retrofitting a variable-speed residential air-source heat pump to run on 350  V DC in both the laboratory and the field. The laboratory testing was conducted for both heating and cooling in a pair of psychrometric chambers that housed the heat pump's indoor and outdoor units, following AHRI Standard 210/240. The field testing was conducted in an occupied test house in West Lafayette, Indiana, during a winter month with outdoor temperatures as low as -8.6~$^\circ$C. To the authors' knowledge, this paper is the first to report laboratory results from retrofitting a variable-speed heat pump to run directly on DC for heating, or to report results from any type of DC heat pump testing in the field.

After reporting laboratory and field test results, this paper uses the field experiment data in system-level modeling of a DC nanogrid. By nanogrid, we mean a kW-scale electrical system that integrates generation, storage, and load within a single building that may be connected to a larger power grid or operate independently. Nanogrids are generally smaller than microgrids, which may reach higher power scales and span multiple buildings. The nanogrid studied here includes 14.3 kW of solar PV, a 20 kWh stationary battery, and the heat pump with 14 kW (4 tons) of nameplate cooling capacity. The model uses one year of measured heat pump load data and rest-of-house load data from the test house, coincident solar irradiance measurements, and battery control logic that prioritizes direct use of solar power by the heat pump. Although the model does not include important effects such as voltage fluctuations, harmonics, stability, grid interaction, protection, or long-term reliability -- important directions for future analysis -- it builds upon existing modeling results by including more real-world load data and more realistic converter part-load efficiency curves. In simulations, the house's annual electricity bills are 12.5\% to 16.7\% lower with a DC nanogrid configuration than with an AC nanogrid configuration.

This paper makes three main contributions to the research literature. First, it presents the first laboratory test results for a DC heat pump retrofit including both heating and cooling tests. Second, this paper presents the first real-world tests of any kind for a DC heat pump retrofit at a field site. Taken together, the laboratory and field test results build confidence that heat pumps can be retrofit to run reliably on DC without significant performance degradation. Third, this paper presents modeling, informed by the first-in-kind field data described above, that estimates the energy efficiency improvements and utility bill savings due to converting a residential solar nanogrid from AC to DC. To the authors’ knowledge, these are the highest fidelity estimates that can be found in the literature. These contributions bring the research community a step closer to trustworthy demonstration of the feasibility and benefits of DC distribution in residential buildings -- key data points for decision-makers in business or government who may consider the case for real-world deployment at scale.

This paper is organized as follows. Section \ref{rel-work} positions the contributions of this paper relative to past research. Section \ref{exp-methods} details the methodology for laboratory and field experiments. Section \ref{exp-results} reports the results of the laboratory and field experiments. Section \ref{sys-mod} presents the nanogrid models and simulation results. Section \ref{discussion} discusses the results, the limitations of this paper, and possible directions for future work. Section \ref{conclusion} concludes the paper.

\section{Related work}
\label{rel-work}

This section discusses DC power systems, focusing on low-voltage residential applications (Section \ref{powerSystems}). Section \ref{equipmentExperiments} then reviews related experimental research on residential DC equipment, focusing on heating and cooling. Section \ref{nanogridModeling} surveys residential DC nanogrid modeling work. Section \ref{ourContributions} positions this paper's contributions relative to past work.

\subsection{DC power systems and the DC Nanogrid House}
\label{powerSystems}

Electric power systems are commonly categorized by voltage as high (typically 50 kV or above), medium (1.5 kV to 50 kV), or low (below 1.5 kV). High-voltage DC transmission has the potential to enable power flows over longer distances with lower losses and higher capacity for a given conductor than high-voltage AC transmission \cite{watson2020overview}, and has begun to see large-scale adoption \cite{wang2022large}. Medium- and low-voltage DC power system technologies are somewhat less mature, but have potential applications in ships, trains, data centers, residential and commercial buildings, electric vehicle charging stations, and some heavy industries \cite{fotopoulou2021state}. The EMerge Alliance was recently created to develop standards for DC and hybrid AC-DC technologies for medium- and low-voltage applications \cite{EMerge-Alliance}.

Medium- and low-voltage DC power systems may be configured as microgrids or nanogrids that can island a community or building, respectively, from the surrounding power system and serve local loads from local generation or storage. Controllable microgrids or nanogrids could improve resilience to extreme weather events and other emergencies by quickly islanding in response to outages in the surrounding power system \cite{shahidehpour_microgrids_2016}. DC microgrids could alleviate issues faced by longer distance distribution of DC power \cite{dastgeer_analyses_2019, dastgeer_comparative_2017}. DC microgrids can also connect DC-powered devices through a common DC bus, increasing the throughput of power from source to sink by removing unnecessary conversions between AC and DC, with modeled energy savings of 10-15\% \cite{dragicevic_dc_2016, chauhan_dc_2018}. Studying DC distribution for a school in Ireland, Alshammari et al. estimated 5\% savings based on measured AC data and assumed converter efficiency curves \cite{alshammari2021dcIreland}. Relative to AC microgrids, DC microgrids could also reduce distribution line losses and provide better voltage regulation \cite{chauhan_dc_2018}.

While DC microgrids and nanogrids could provide a range of benefits, they face several barriers to real-world adoption. DC distribution systems for residential and commercial buildings face a lack of cohesive standards and high capital costs \cite{elsayed_dc_2015, fotopoulou2021state}. Protection for DC distribution systems in occupied buildings also presents challenges, as safely and reliably breaking an active DC circuit can be difficult \cite{fotopoulou2021state}, although recent innovations such as fault-managed power \cite{lorusso2025introduction} might address protection issues. Installing and maintaining DC distribution systems can also entail higher costs and longer delays than AC systems, as electricians may have less experience with DC and appropriate DC equipment may not be commercially available \cite{fotopoulou2021state}.

\begin{figure}
  \centering
  \includegraphics[width=0.95\columnwidth]{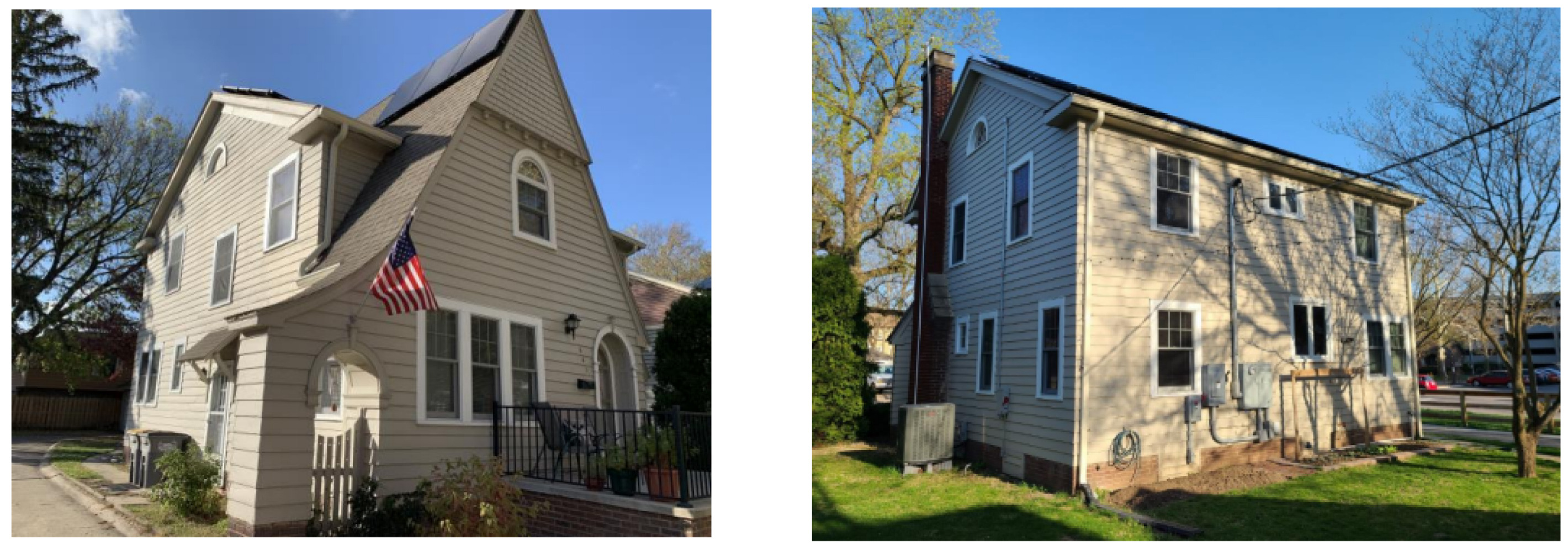}
  \caption{The DC Nanogrid House is a 1920s-era, 208 m$^2$, two-story, detached single-family house in West Lafayette, Indiana.}
  \label{fig:dc-house}
\end{figure}

Toward addressing some of these barriers to adoption, a team of researchers at Purdue University has renovated a test house with DC distribution and devices \cite{ore_analysis_2020, ore_case_2021, ore-thesis-2021}. The DC Nanogrid House, shown in Figure \ref{fig:dc-house}, is a 1920s-era, 208 m$^2$, two-story detached single-family house occupied by three graduate student researchers. The house has been renovated with insulation, new windows, all-electric appliances, a 14.3 kW solar photovoltaic array, a 20 kWh stationary battery, and a custom home energy management system \cite{pergantis2024field, pergantis2024humidity, pergantis2025protecting}. The house has both conventional AC wiring and a DC distribution system that is under ongoing development. After the retrofits described in this paper, the central variable-speed heat pump can now be switched between the AC and DC distribution systems through a lockout/tagout system. The project aims to retrofit additional devices -- including an electric vehicle charging station, a full appliance suite, and lighting -- to run on DC, with the ultimate goal of characterizing the performance differences between running a renovated 1920s house on DC vs. AC. Retrofitting and testing the heat pump is an important step toward this long-term goal: The heat pump uses over two-thirds of annual energy in the house, which is located in a climate that sees both hot, humid, sunny summers and very cold winters.

\subsection{Residential DC appliance experiments}
\label{equipmentExperiments}

Many works that discuss efficiency differences between AC and DC appliances only do so with modeling. The few experimental works of which the authors are aware considered small-scale appliances that operate at 48  V DC (\cite{chacko_dc_2022, sriram_development_2024}). The only previous experimental publication that compared an at-scale test of a residential heat pump was also done in psychrometric chambers with the same heat pump evaluated in this paper \cite{ore_evaluation_2020}. That experiment was limited to cooling tests and did not evaluate the variable-speed functionality of the heat pump. The focus in \cite{ore_evaluation_2020} on cooling was likely due to the facts that air conditioning is much more prevalent than electric heating in North American housing, and that cooling loads align better with solar PV output, which peaks in the summer months. The heat pump was tested according to AHRI Standard 210/240 conditions, but the setpoint of the heat pump was chosen such that the unit ran at full speed across the tests but was not explicitly controlled. Under these conditions, there was a 6-8\% increase in system COP with DC power. Apart from Ore et al.  \cite{ore_evaluation_2020}, the authors are aware of no papers that directly compared AC and DC power inputs for a variable-speed residential heat pump. 

In \cite{vossos_adoption_2022}, Vossos et al.  generated a map, shown in Figure \ref{fig:DC-US-2022}, of DC projects (both small-scale demonstrations and full systems) in the United States and Canada as of 2022. Nearly all projects focus on lighting. The only heating or cooling system listed is the one tested by Ore et al.  \cite{ore_evaluation_2020}.

{ Beyond heat pumps, several studies have experimentally investigated efficiency improvements from powering smaller appliances on DC. Chacko et al. compared light-emitting diode bulbs and fans between AC and DC power, finding 20 to 40\% energy savings from switching all lights and fans to DC \cite{chacko_dc_2022}. Fregosi et al.  \cite{fregosi_comparative_2015} demonstrated 6-8\% higher on-site PV consumption from powering light-emitting diode bulbs on DC. Fregosi et al. also modeled multiple building types across the United States, finding that increases in PV self-consumption were larger in sunnier states. Sriram et al. compared efficiency differences between powering light-emitting diode bulbs on AC vs. DC power-over-ethernet \cite{sriram_development_2024} in the DC Nanogrid House. They found that for small numbers of bulbs, fixed loads from power-over-ethernet networking equipment outweighed energy savings. When scaled up to all the bulbs in a typical house, however, savings more than offset networking loads, with net savings reaching up to 22\%.}

\subsection{Residential DC nanogrid modeling}
\label{nanogridModeling}

The device-level experiments surveyed above help build confidence that individual loads can function reliably on DC with equivalent or better performance than on AC. However, device-level experiments are insufficient to characterize the system-level benefits of connecting multiple DC devices through DC distribution. While system-level benefits could in principle be evaluated in hardware, it is challenging today to find appropriate DC distribution components (such as DC/DC converters and control systems) or loads that can connect directly to DC distribution without after-market retrofits. System-level modeling, while less realistic than field experiments, can provide preliminary estimates of system-level benefits and can explore implementation challenges and solutions.

Several studies have modeled the system-level benefits, such as reduced conversion losses, of DC nanogrids in residential buildings with rooftop solar PV \cite{kakigano_loss_2010, muruganantham_challenges_2017}. Frances et al. presented a range of options for modeling the efficiencies of different DC-DC, AC-DC, and DC-AC conversions \cite{frances_modeling_2018}. Vossos et al. evaluated several residential microgrid or nanogrid architectures and applied simple efficiency calculations \cite{vossos_energy_2014}. In follow-up work, Vossos et al. compared the nameplate efficiencies of commercially available products that offer both AC and DC options, showing that DC counterparts were often 20-30\% more efficient \cite{vossos-2017-DC-appliances}. Garbesi et al. estimated that a DC configuration can save 7 or 13\% energy with or without energy storage, respectively \cite{garbesi_optimizing_nodate}. Vossos et al. \cite{vossos_energy_2014} and Garbesi et al. \cite{garbesi_optimizing_nodate} both modeled space conditioning loads only for cooling; neither explored powering electric heating equipment on DC. Fapi et al. developed models to explore DC microgrid control with a PV array, a wind turbine, and a heat pump \cite{nzoundja_fapi_control_2025}. The PV array and wind turbine were modeled in detail, while the less detailed heat pump model assumed the coefficient of performance (COP) was a linear function of the outdoor temperature. Although Fapi et al. modeled a heat pump, they did not evaluate the energy savings from switching the heat pump from AC to DC. {Table \ref{ta:nanogrid-literature} summarizes these studies.}

\begin{table}
\centering
\caption{Summary of closely related studies on DC residential systems}
\vspace*{6pt}

\begin{tabular}{@{}p{1.0cm}p{0.8cm}p{2cm}p{3.8cm}p{4.5cm}@{}}
      Study & Year & Experiments? & Scope & Main results \\
      \midrule
      \cite{ore_evaluation_2020} & 2020 & Yes & Heat pump in cooling mode & 6--8\% COP improvement from DC \\
      \cite{sriram_development_2024} & 2024 & Yes & Power-over-ethernet light-emitting diodes & $\leq$20\% lighting energy savings  from DC \\
      \cite{chacko_dc_2022} & 2022 & Yes & Light-emitting diodes, brushless DC fan motors & 20--40\% energy savings from DC \\
      \cite{vossos_energy_2014, garbesi_optimizing_nodate} & 2014 & No & PV and house loads in 14 cities & 5\% energy savings from DC without battery, 14\% with \\
      \cite{kakigano_loss_2010} & 2010 & No & PV and appliances & 15\% lower energy losses with DC\\
      \cite{vossos-2017-DC-appliances} & 2017 & No & Market analysis & Few market-ready DC components \\
      \cite{nzoundja_fapi_control_2025} & 2025 & No  & Heat pump, battery, PV, wind & $\leq$30\% energy savings from predictive control \\
\end{tabular}
\label{ta:nanogrid-literature}
\end{table}

\subsection{Contributions of this paper}
\label{ourContributions}

In summary, this paper advances the state of knowledge on residential DC equipment and nanogrids in two main ways. First, this paper presents the first field demonstration that any major residential appliance -- in this case, a heat pump, one of the largest energy users in electrified North American housing -- can be readily retrofit and reliably operated on DC power. Expanding on prior work with the same heat pump \cite{ore_evaluation_2020}, this paper also presents the first DC heat pump laboratory experiment results that include heating and variable-speed operation (rather than running steadily at maximum capacity). The laboratory and field experiments suggest that performance differences at the device level between AC and DC operation are {statistically insignificant}.

Second, this paper uses field data to model the system-level benefits of incorporating the tested heat pump into a DC solar nanogrid. While modeling a residential DC solar nanogrid is not new, to the authors' knowledge, this paper a) is the first to use real field data from a major appliance in a calibrated model, rather than relying on assumed performance parameters, and b) is the first to model DC heat pump operation year-round, including during peak heating conditions. The modeling suggests that, relative to an AC configuration, the DC configuration could {reduce annual electricity bills by 12.5\% to 16.7\%. This is in line with previous research results, which typically show 10\% to 20\% energy savings \cite{vossos_energy_2014, garbesi_optimizing_nodate, kakigano_loss_2010}.}

\section{Experimental methodology}
\label{exp-methods}

\begin{figure}
  \centering
  \includegraphics[width=0.99\columnwidth]{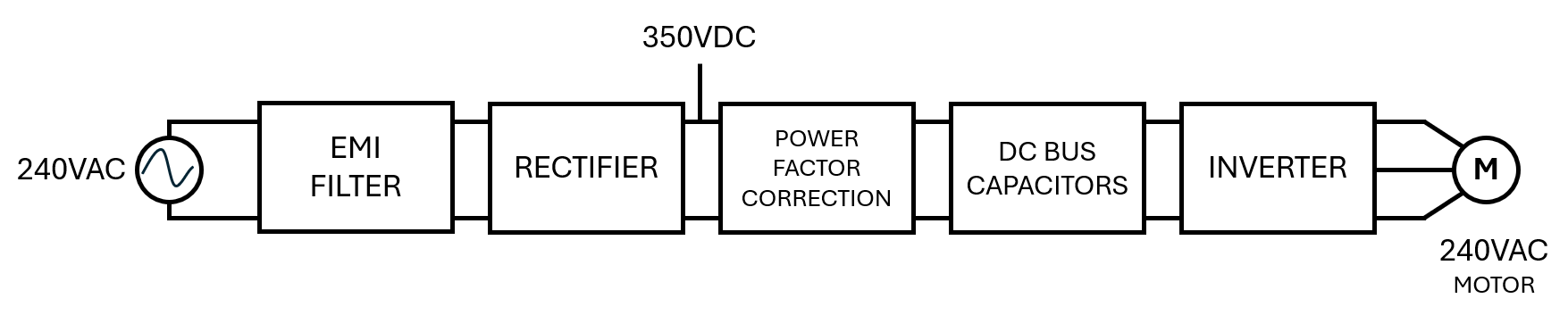}
  \caption{Simplified heat pump compressor drive showing the main electrical blocks of the system. EMI stands for electromagnetic interference.}
  \label{fig:Trane-drive}
\end{figure}

An off-the-shelf variable-speed residential split-system heat pump with a scroll compressor and 14 kW (4 tons) of nameplate cooling capacity was retrofitted to accept both AC and DC inputs. Figure \ref{fig:Trane-drive} shows the unit's power drive. For this study, {\color{black} the 240 V AC input was replaced with a 350 V DC supply}. The unit was charged with 6.1 kg of R-410A to achieve 5 K of subcooling under the AHRI Standard 210/240 A2 conditions (Table \ref{ta:AHRI-210-240}) as per the manufacturer’s specifications. {To match the subcooling specification, the heat pump was run in steady state at the A2 test point (indoor chamber air: 27 $^\circ$C temperature, 51\% relative humidity; outdoor chamber air: 35 $^\circ$C, 40\%) with an initial, uncalibrated mass of its refrigerant (R-410A). Incremental refrigerant was then gradually added until the measured refrigerant temperature at the outlet of the condenser coil was 5 K cooler than the saturation temperature of the refrigerant at the measured condenser outlet pressure. The same subcooling calibration procedure was conducted before AC testing and before DC testing, allowing direct performance comparison.} 

\begin{table}
  \caption{Test conditions from AHRI Standard 210/240}
  \vspace*{6pt}
  \centering
    \begin{tabular}{@{}ccccc@{}}
               & \bf Indoor & \bf Outdoor & \bf Indoor & \bf Outdoor \\
      \bf Test & \bf dry bulb & \bf dry bulb & \bf RH & \bf RH\\
               & \bf [$^\circ$C] & \bf [$^\circ$C] & \bf [\%] & \bf [\%] \\
      \midrule
      A2 & 27 & 35 & 51 & 40 \\
      B2 & 27 & 28 & 51 & 40 \\
      B1 & 27 & 28 & 51 & 20 \\
      Ev & 27 & 31 & 51 & 20 \\
      F & 27 & 19 & 51 & 40 \\
      H01 & 21 & 17 & 56 & 72 \\
      H11 & 21 & 8 & 56 & 72 \\
      H1n & 21 & 8 & 56 & 73 \\
      H2v & 21 & 2 & 56 & 82 \\
      H32 & 21 & -8 & 56 & 70 \\
    \end{tabular}
  \label{ta:AHRI-210-240}
\end{table}

Laboratory tests were performed in a pair of psychrometric chambers following AHRI Standard 210/240 \cite{ahri-210-240}. For AC testing, power was supplied to the heat pump at a nominal split-phase 230  V AC. The DC configuration operated at a nominal bi-polar 350  V DC input. {For both AC and DC testing, the auxiliary heating element was not present.} Thermal and electrical performance was evaluated under variable-speed operation using the AHRI Standard 210/240 test procedures as shown in Table \ref{ta:AHRI-210-240}. To verify the testing procedures, an energy balance was calculated between the indoor unit refrigerant coil and the air blown across it to within 6\% (Figures \ref{fig:AC-energy-bal} and \ref{fig:DC-energy-bal}). 

Field tests were conducted at the DC Nanogrid House in West Lafayette, Indiana \cite{ore-thesis-2021}. This is a ``living laboratory'' with three graduate student occupants. The house provides realistic boundary conditions for dynamic testing. A heat pump of the same make and model studied in the laboratory was installed. The refrigerant charge was calibrated to provide 5 K of subcooling in the A2 test condition. From the manufacturer recommendation, only the outdoor unit was retrofitted to run on DC. {The outdoor unit (compressor, heat exchanger, and fan) uses more than 90\% of the heat pump's overall energy. The indoor unit consists of heat exchanger, a fan for air circulation, and auxiliary electric resistance heating elements (large resistors) controlled by AC relays. The manufacturer stressed that attempting to power these AC relays on 350 V DC could present a fire risk, so the indoor unit was left on AC power. Future studies should investigate methods to retrofit heating element controls to run safely on DC. However, powering controlled heating elements on DC vs. AC should not change their energy use significantly. A DC retrofit should only affect the controls, which typically draw less than 1 W. Regardless of whether the heating elements themselves are powered on AC or DC, they should draw 5--20 kW for Joule heating.}

\subsection{Steady-state laboratory test descriptions}

The heat pump was placed in a pair of psychrometric chambers to simulate indoor and outdoor conditions using AHRI Standard 210/240. Table \ref{ta:AHRI-210-240} shows the steady-state test conditions for variable-speed unitary split residential systems.

\begin{figure}
  \centering
  \includegraphics[width=0.85\columnwidth]{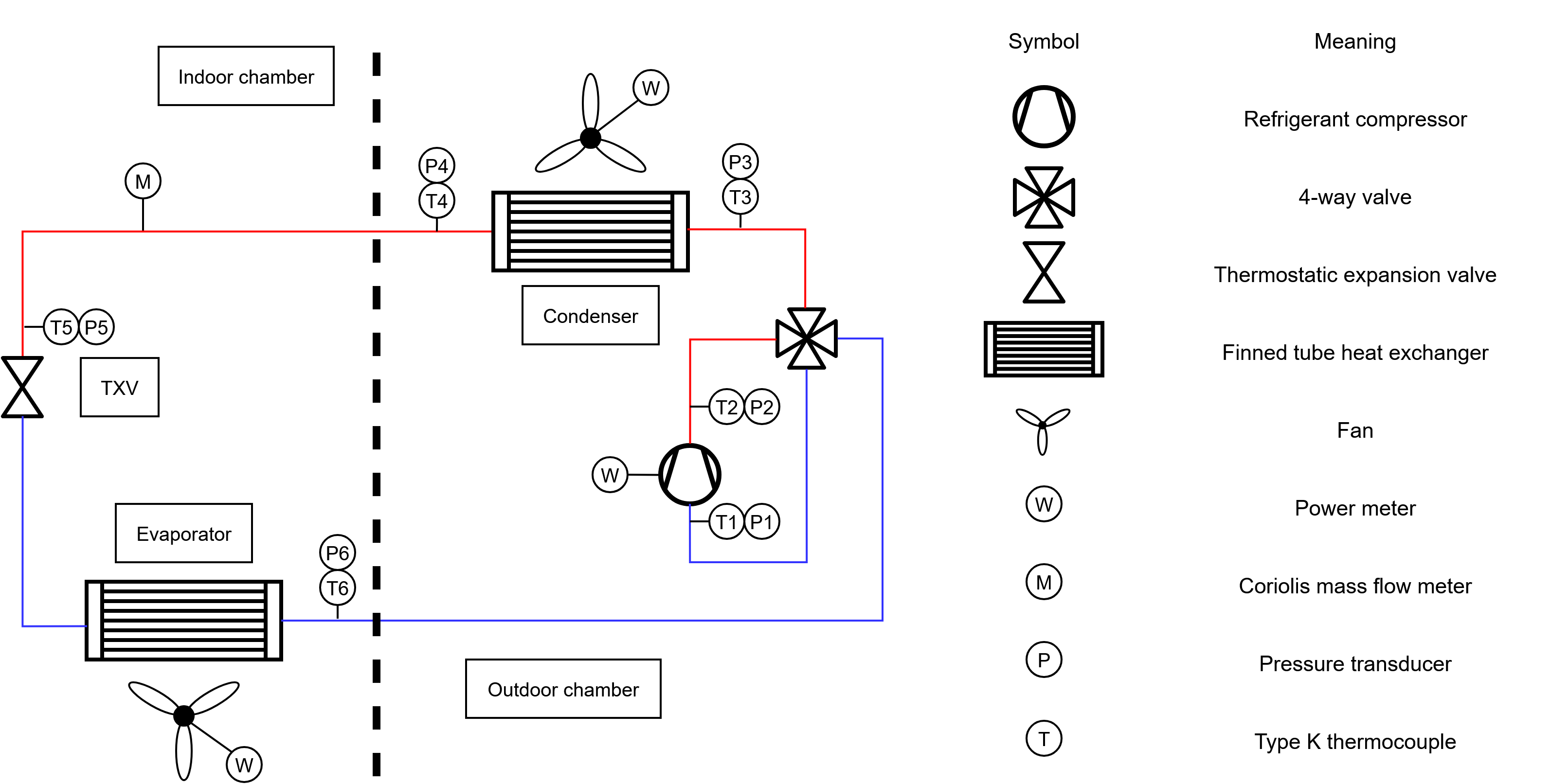}
  \caption{Piping and instrumentation diagram of the heat pump cycle in cooling mode.}
  \label{fig:PID_DCHP_cool}
\end{figure}

The heat pump was instrumented with thermocouples and pressure transducers to directly measure the refrigerant and air properties. Laboratory data was collected using a National Instruments cRIO-2091 data acquisition system and the communication and data handling systems described in \cite{pergantis2023sensors}. Figure \ref{fig:PID_DCHP_cool} shows a piping and instrumentation diagram (adapted from \cite{ore_evaluation_2020}), which details the placement  for cooling mode of the thermocouples, pressure transducers, and a Coriolis mass flow meter in the refrigerant cycle. The four-way valve in Figure \ref{fig:PID_DCHP_cool} switches the heat pump to heating mode by reversing the direction of refrigerant flow so that the indoor and outdoor heat exchangers function as the condenser and evaporator, respectively. Table \ref{ta:state-points} shows the refrigerant states, which are numbered the same for heating and cooling. 

Table \ref{ta:uncertainty} shows types of sensors used and the associated measurement uncertainty ranges. Air dry-bulb temperature data was collected using three-by-three thermocouple grids at the inlet and outlet of the indoor unit. A chilled mirror dewpoint temperature humidity sensor was used to calculate the relative humidity at the outlet of the indoor unit. A capacitive relative humidity sensor was placed at the inlet to the indoor unit. The dry-bulb temperature and relative humidity were used to calculate an energy balance between the air and refrigerant as specified in AHRI Standard 210/240. Since the system is variable-speed, the manufacturer provided a proprietary testing controller to operate the unit under the AHRI Standard 210/240 test conditions. This controller adjusts the revolutions per minute of the outdoor fan, indoor fan fan, and compressor. 

\begin{table}
  \caption{State descriptions and measurements}
  \vspace*{6pt}
  \centering
    \begin{tabular}{@{}clc@{}}
      \bf State & \bf Description & \bf Measurement \\
      \midrule
      1 & Compressor suction & T,P \\
      2 & Compressor discharge & T,P \\
      3 & Condenser inlet & T,P \\
      4 & Condenser outlet & T,P \\
      5 & TXV inlet & T,P \\
      6 & TXV outlet & None \\
      7 & Evaporator inlet & None \\
      8 & Evaporator outlet & T,P \\
    \end{tabular}
  \label{ta:state-points}
\end{table}

\begin{table}
  \caption{Measurement uncertainty in the steady-state testing}
  \vspace*{6pt}
  \centering
  
    \begin{tabular}{@{}l l l l l@{}}
        \textbf{Measurement} & \textbf{Uncertainty} & \textbf{Range} & \textbf{Manufacturer} \\
        \midrule
        Temperature & $\pm 0.5$ $^\circ$C & $[-200, 1260]$~$^\circ$C & Type-K thermocouple \\
        Pressure & $\pm 5$ kPa & $[0, 7000]$ kPa & Dwyer Omega \\
        Mass flow rate & $\pm 0.01$ g/s & $[0, 600]$ g/s & Emerson \\
        Indoor unit total power & $\pm 5$ W & $[0, 1000]$ W & Ohio Semitronics \\
        Outdoor unit compressor power & $\pm 12.5$ W & $[0, 5000]$ W & Ohio Semitronics \\
        Outdoor unit fan power & $\pm 5$ W & $[0, 1000]$ W & Ohio Semitronics \\
    \end{tabular}
  \label{ta:uncertainty}
\end{table}

Two sets of Watt transducers were required to gather data on the AC and DC configurations. The key difference between a DC and AC Watt transducer is the current measurement method, which is either through a Hall effect sensor (DC) or an inductive measurement (AC). For the AC configuration, three inductive Watt transducers (Ohio Semitronics model number GW5020D) were used to collect data from the total power to the indoor unit, the outdoor fan, and the compressor. The DC configuration used two Hall-effect Watt transducers to separately measure the power drawn by the indoor unit and the outdoor unit. The uncertainty of the power meters for the indoor fan and the outdoor fan were $\pm$5 W at full scale, while the outdoor unit power consumption was rated at $\pm$12.5 W at full scale.

\subsection{Dynamic field test descriptions}

\begin{figure}
  \centering
  \includegraphics[width=0.95\columnwidth]{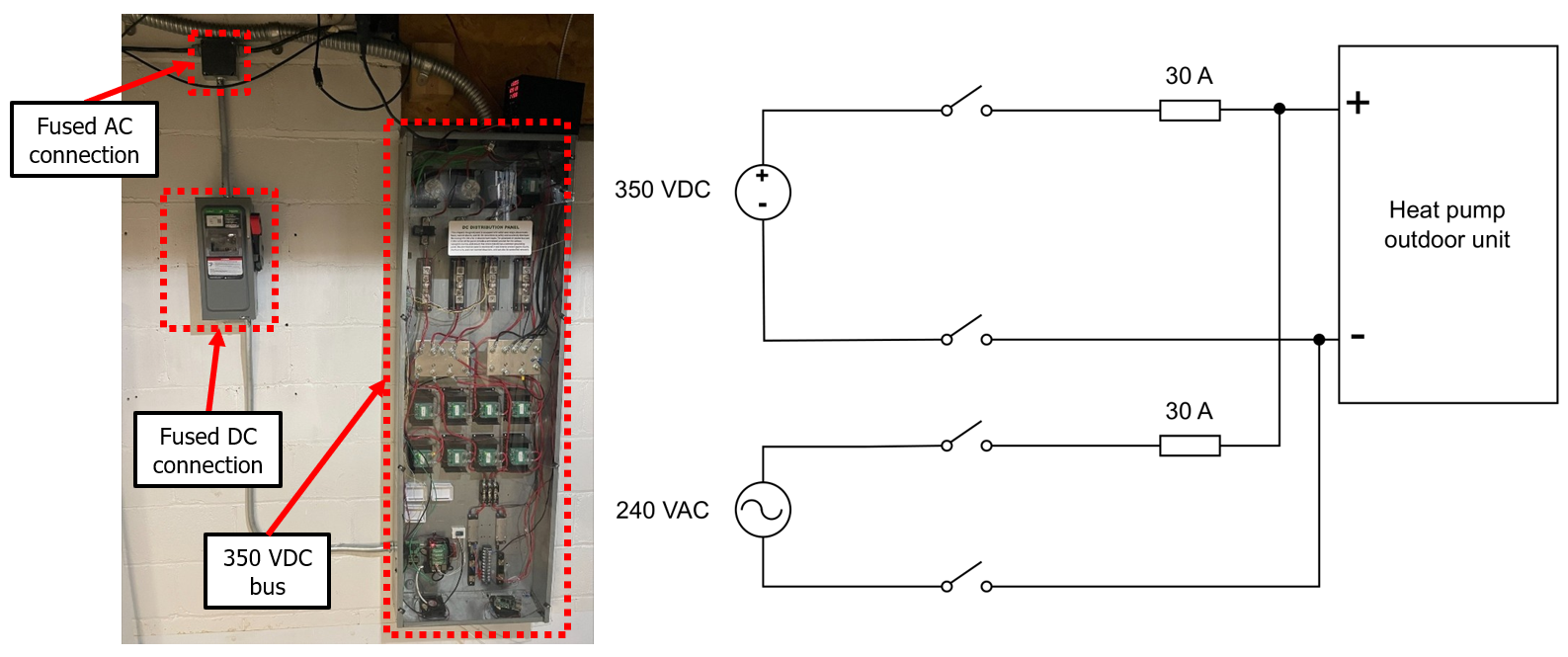}
  \caption{Wiring of the outdoor unit to the 350 VDC bus at the DC Nanogrid House using a lockout/tagout system.}
  \label{fig:loto}
\end{figure}

A heat pump of the same make and model tested in the psychrometric chambers was installed at the DC Nanogrid House in West Lafayette, Indiana. {West Lafayette is in climate zone 5A, which typically sees 3,000 to 4,000 heating degree-days per year (calculated using a balance-point temperature of 18 $^\circ$C) \cite{ashrae-climates}.} The heat pump operated on a standard AC input from 2019 to 2024, when it was retrofitted to enable operation on AC or DC.  A lockout/tagout system was implemented to enable manual switching between AC and DC. Figure \ref{fig:loto} shows the physical installation and accompanying wiring diagram.

The heat pump was run with a constant indoor air temperature setpoint of 20.5 $^\circ$C from December 18, 2024, through January 16, 2025. The power was measured from the on-board power transducers on the DC nanogrid's bi-directional inverter. The on-board power transducers are accurate to within $\pm$5 W. Historical AC data was directly measured with sub-metering on the breaker panel with sensors that are accurate to within $\pm$5 W. 

{The AC field data were measured one year prior to the DC field data, which could lead to somewhat higher performance estimates for AC than DC. Even under standard AC operation, heat pump performance can degrade from one year to the next due to machine wear and tear. Discussions with the manufacturer also suggested that the after-market DC retrofit was not ideal for this heat pump because the compressor's variable-speed drive was not designed to natively accept DC. Operating the compressor drive on DC could therefore accelerate performance degradation relative to AC operation.} 

\begin{figure}
  \centering
  \includegraphics[width=0.75\columnwidth]{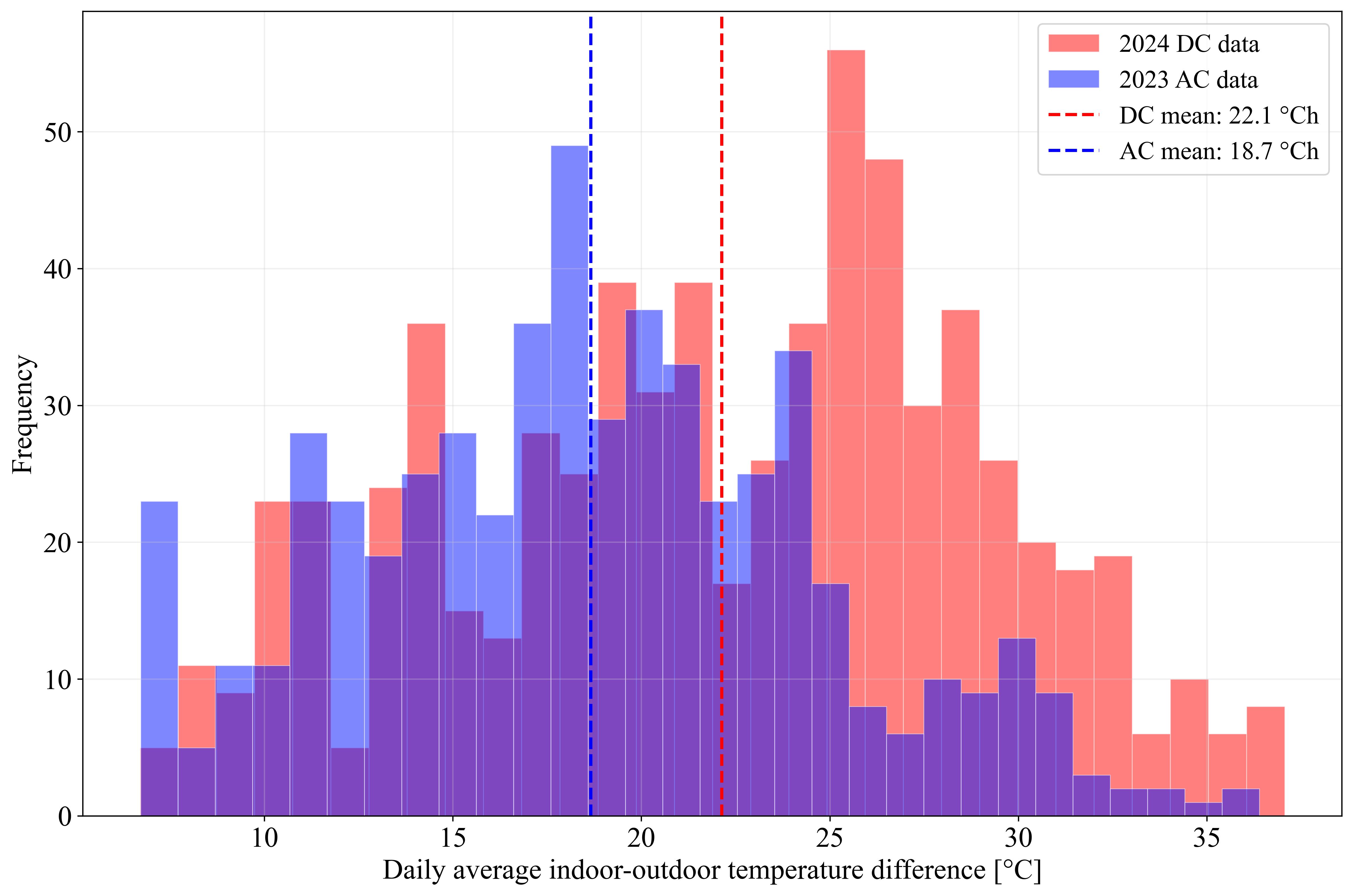}
  \caption{Histograms of daily average indoor-outdoor temperature difference for DC and AC testing. On average, AC field testing was conducted in somewhat warmer weather, which could lead to somewhat higher performance estimates.}
  \label{fig:HDH-freq-analysis}
\end{figure}

{The AC field data were selected from periods when the thermostat setpoint matched the 20.5 $^\circ$C used for DC testing, and when weather conditions approximately matched the DC weather conditions. Figure \ref{fig:HDH-freq-analysis} shows histograms of the daily average indoor-outdoor temperature difference for each day of the DC data (red) and the AC data (blue). More precisely, Figure \ref{fig:HDH-freq-analysis} shows histograms of 
\begin{equation}\label{eq:HDH}
    \frac{ 1 }{K} \sum_{k=1}^K \max(0, 20.5 \ ^\circ\text{C} - T_\text{out}(k)) ,
\end{equation}
where the integer $k$ indexes time steps, $K$ is the number of time steps per day, 20.5 $^\circ$C is the indoor air temperature setpoint, and $T_\text{out}(k)$ ($^\circ$C) is the outdoor air temperature. The $\max(0,\cdot)$ term in Equation \eqref{eq:HDH} eliminates times when it is warmer outside than inside and heat demand is expected to be zero. While the AC and DC datasets in Figure \ref{fig:HDH-freq-analysis} spanned nearly identical temperature ranges, the DC mean (dashed blue line) of 21.1 $^\circ$C was 3.4 $^\circ$C higher than the AC mean (dashed red line) of 18.7 $^\circ$C. This indicates that the AC testing weather was somewhat warmer on average than the DC testing weather, which could lead to somewhat higher performance estimates for AC than DC.}

\section{Experimental results}
\label{exp-results}

\subsection{Steady-state laboratory test results}

Table \ref{ta:AC-DC-results-210-240} shows the heat pump's steady-state laboratory performance data on AC and DC, respectively. This table includes the test condition (from Table \ref{ta:AHRI-210-240}), thermal capacity delivered by the unit, indoor and outdoor power consumption, total power consumption, and system COP. The system COP was calculated as 
\begin{equation}\label{eq:COP-calc}
    \text{COP} = \frac{ \lvert \dot{Q}_\text{indoor} + \dot{Q}_\text{in,fan} \rvert }{\dot{W}_\text{in,fan} + \dot{W}_\text{out,fan} + \dot{W}_\text{comp}} , 
\end{equation}
{where $\dot{Q}_\text{indoor}$ (kW) denotes the rate of heat transfer from the refrigerant to the indoor air (positive in heating mode, negative in cooling mode), $\dot{Q}_\text{in,fan}$ (kW) denotes the rate of heat transfer from the indoor fan to the indoor air (positive in both heating and cooling modes; assumed equal to $\dot{W}_\text{in,fan}$), $\dot{W}_\text{in,fan}$ (kW) denotes the indoor fan input power, $\dot{W}_\text{out,fan}$ (kW) denotes the outdoor fan input power, and $\dot{W}_\text{comp}$ (kW) denotes the compressor input power. The $\dot W$ terms are all positive in both heating and cooling modes.}

\begin{figure}
  \centering
  \subfloat{\label{fig:AC-energy-bal}\includegraphics[width=0.42\columnwidth]{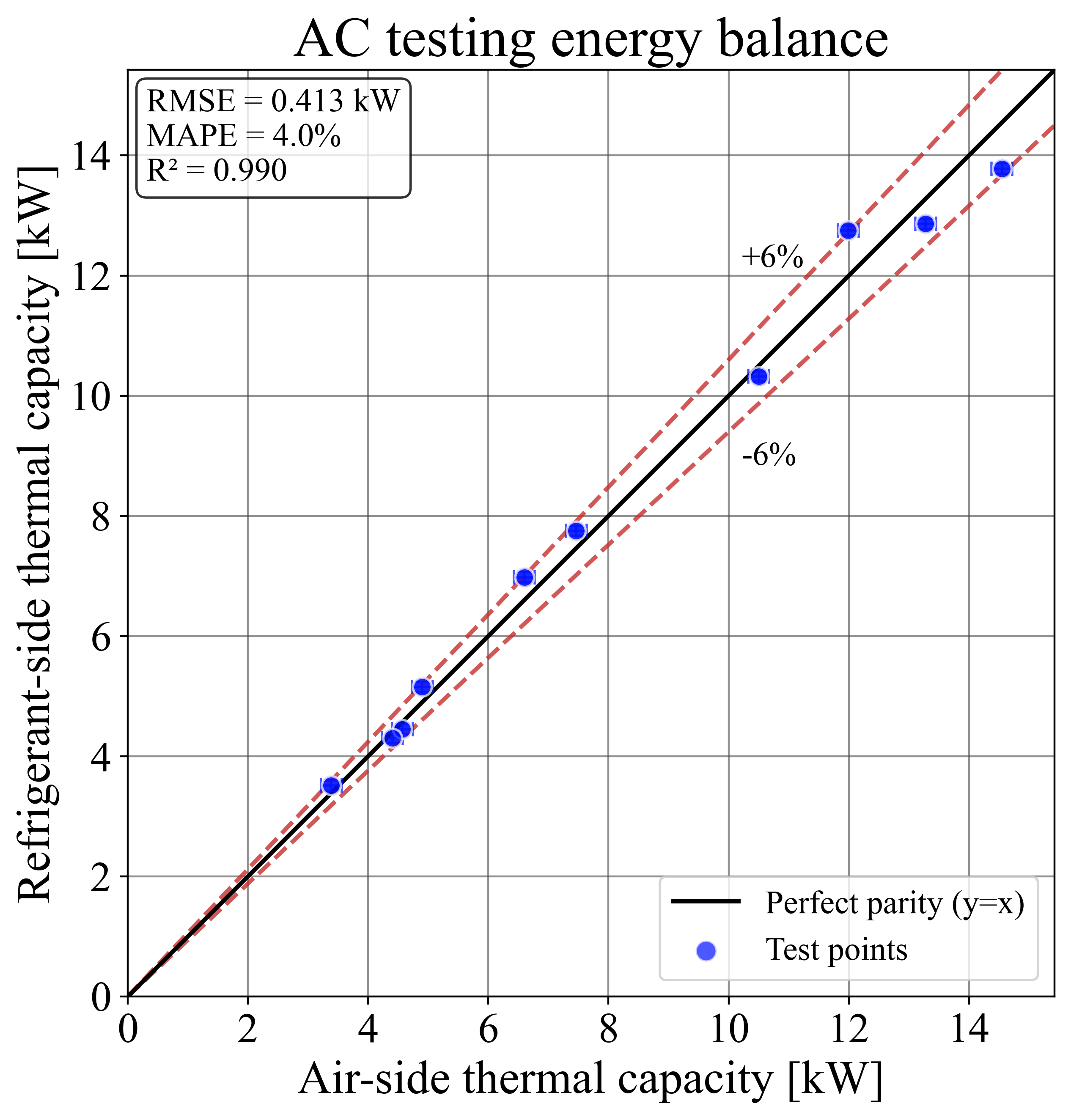}}
  \hfill
  \subfloat{\label{fig:DC-energy-bal}\includegraphics[width=0.42\columnwidth]{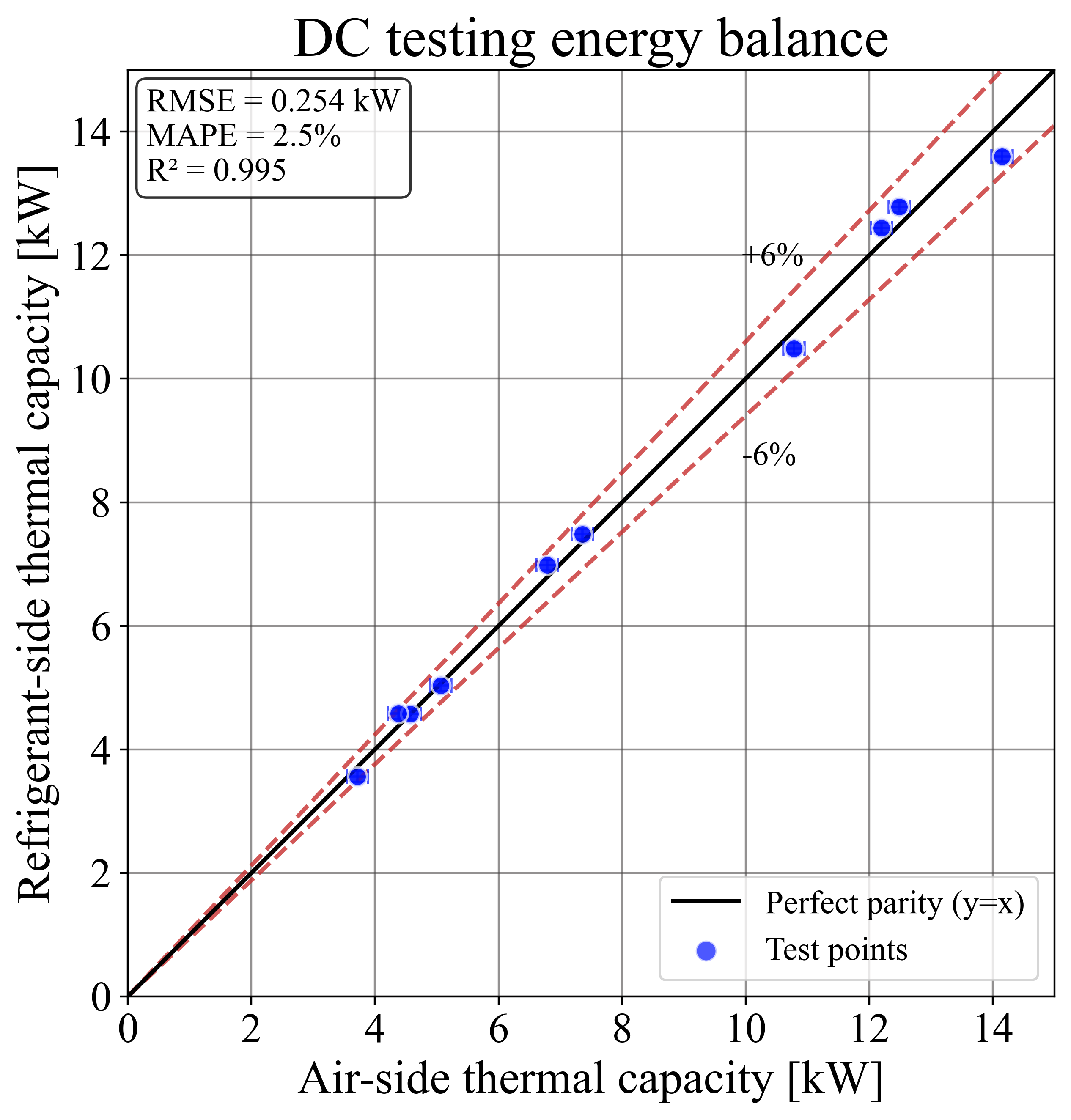}}
  \caption{Air- and refrigerant-side thermal capacity measurements agreed to within $\pm$6\% error on AC (left) and DC (right).}
  \label{fig:energy-balance}
\end{figure}

Figure \ref{fig:energy-balance} shows that refrigerant-side and air-side thermal capacity measurements agreed to within $\pm$6\% error under all test conditions on both AC and DC, as required by AHRI Standard 210/240. The AC test data had a mean absolute percent error of 4\% between the refrigerant-side and air-side measurements. The DC test data had a mean absolute percent error of 2.5\%. This builds confidence that thermal capacity was measured accurately.

\begin{figure}
  \centering
  \includegraphics[width=0.5\columnwidth]{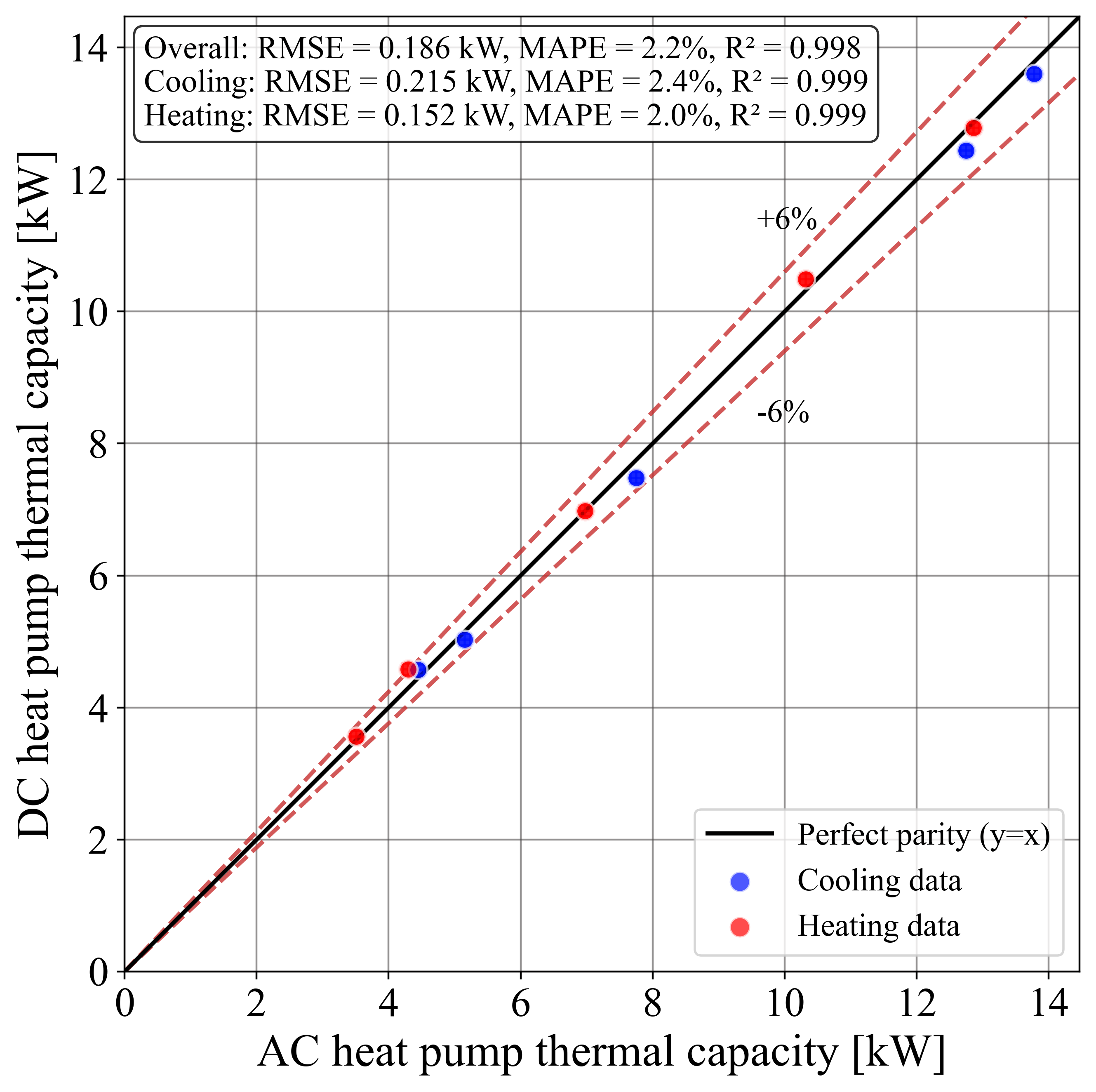}
  \caption{AC and DC thermal capacity measurements agreed to within $\pm$6\% for both heating (red) and cooling (blue).}
  \label{fig:thermal-parity}
\end{figure}

Figure \ref{fig:thermal-parity} shows that the thermal capacity measurements on DC agreed with the AC measurements to within $\pm$6\% error under all test conditions. The mean absolute percent error between AC and DC thermal capacity was 2.4\% for cooling and 2.0\% for heating. This suggests that the heat pump provides the same thermal capacity on DC as on AC.

\begin{figure}
  \centering
  \includegraphics[width=0.5\columnwidth]{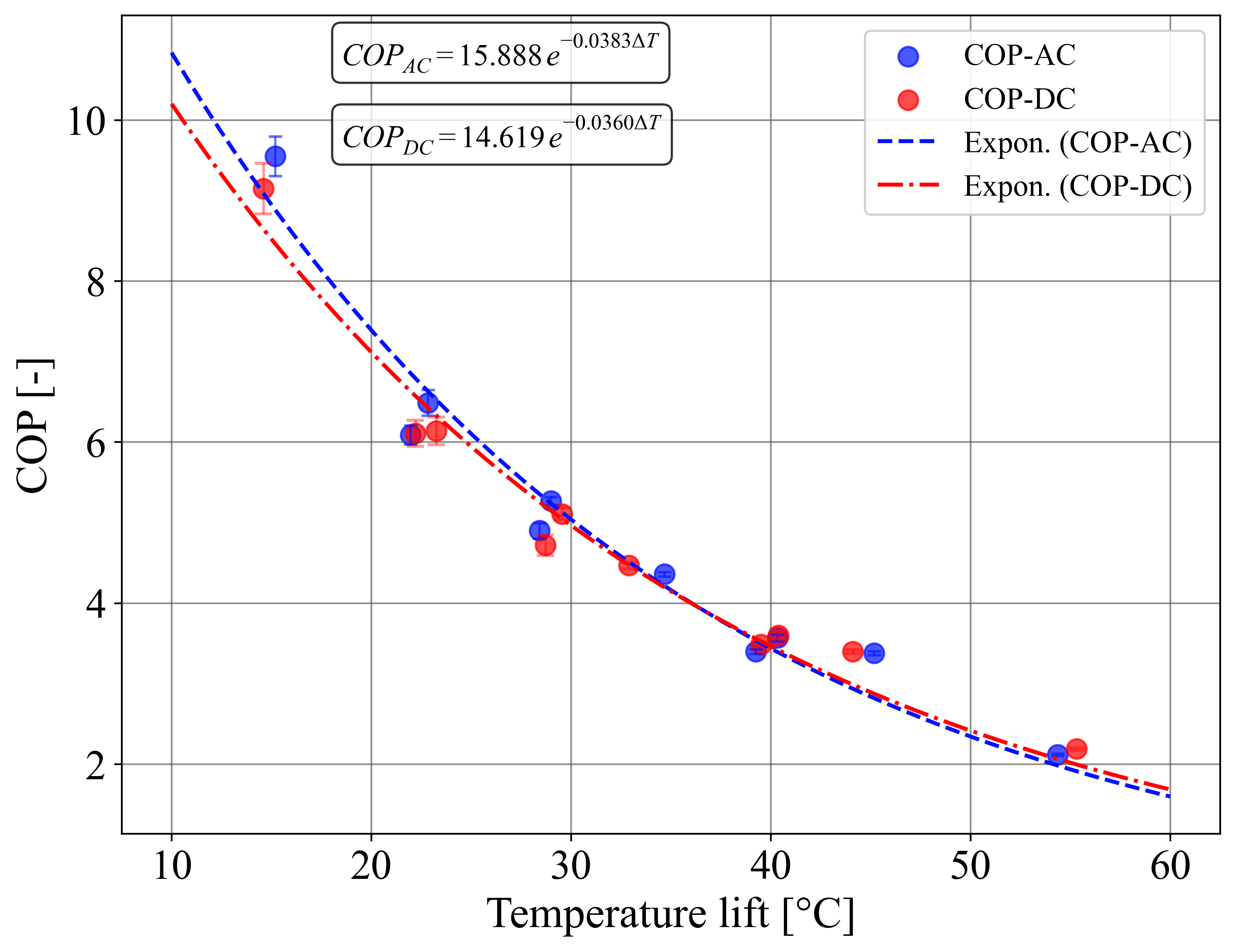}
  \caption{COP as a function of the temperature lift across all steady-state test conditions.}
  \label{fig:COP-temp-lift}
\end{figure}

Figure \ref{fig:COP-temp-lift} shows the AC and DC COP measurements vs. the temperature lift (meaning the difference between the measured condensing and evaporating temperatures for the refrigerant). {The heat pump performed slightly worse (3--4\% lower COP) on DC than on AC in low-load conditions, when the temperature lift was lower. By contrast, the heat pump performed slightly better (1--3\% higher COP) on DC than on AC during high-load conditions, when the temperature lift was higher. However, the relative measurement uncertainty was larger in low-load conditions because the absolute measurement uncertainty ranges were independent of the measurement magnitudes. For example, the outdoor unit power meter was accurate to within $\pm$12.5 W. This is 0.25\% relative uncertainty at a high-load condition where the outdoor unit draws 5 kW, but 4.2\% relative uncertainty at a low-load condition where the outdoor unit draws 300 W. The error bars in Figure \ref{fig:COP-temp-lift}, which were calculated from the measurement uncertainty ranges in Table \ref{ta:uncertainty}, are therefore more pronounced at low temperature lifts than at high temperature lifts.}

\subsection{Dynamic field test results}

\begin{figure}
  \centering
  \includegraphics[width=0.475\columnwidth]{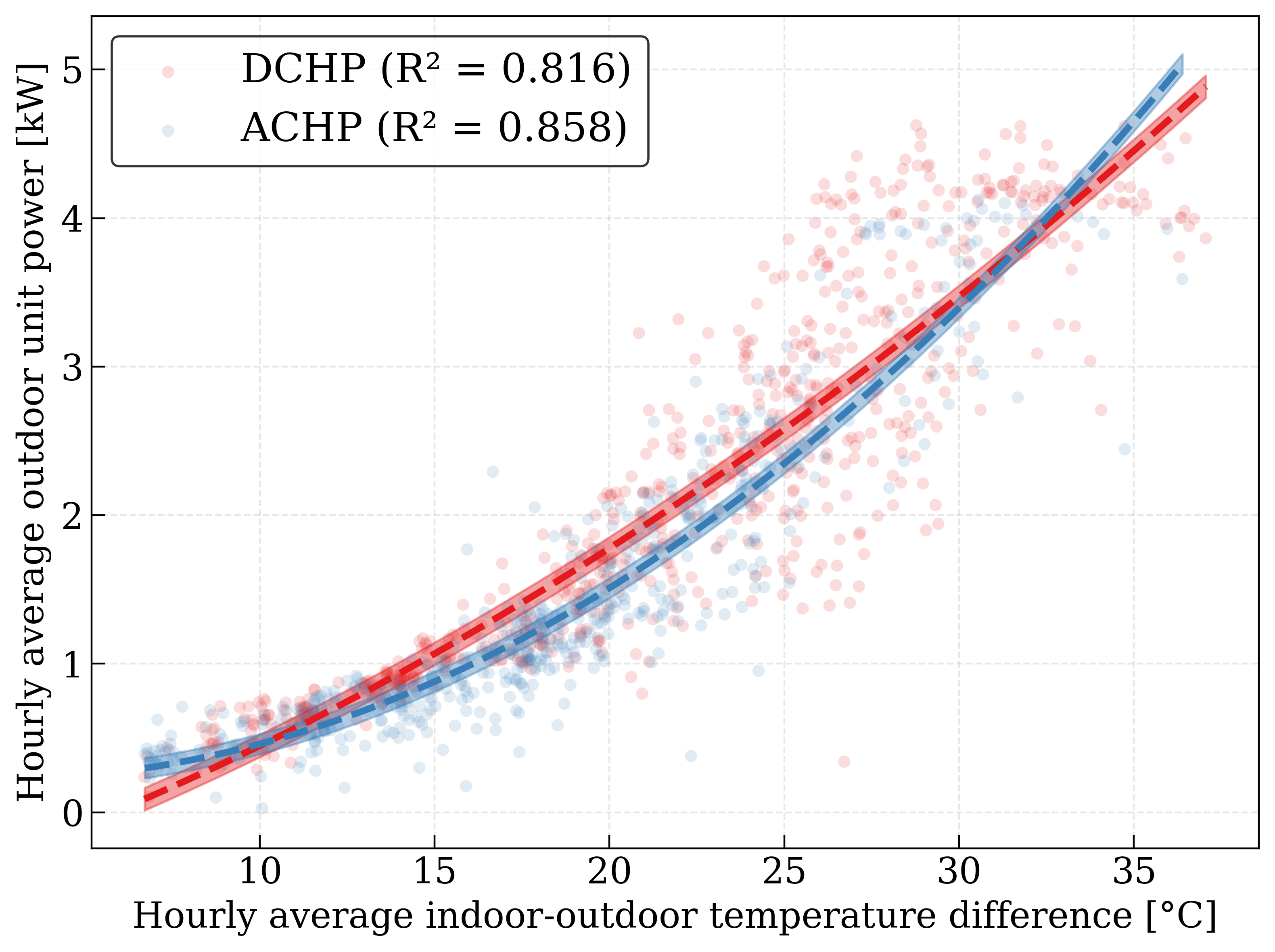}
  \includegraphics[width=0.475\columnwidth]{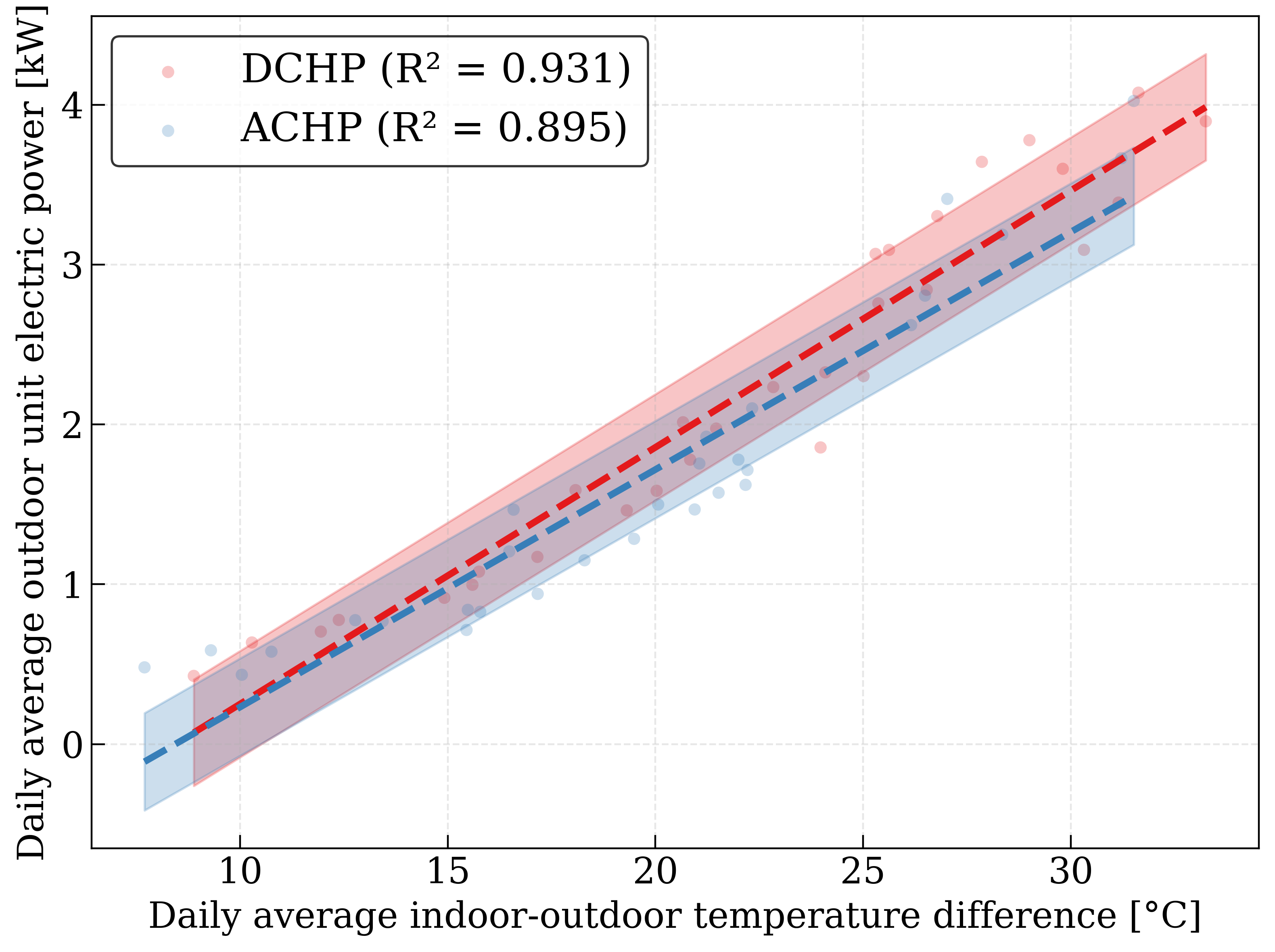}
  \caption{Hourly (left) and daily (right) average outdoor unit input power vs. average indoor-outdoor temperature difference.}
  \label{fig:power-vs-temperature}
\end{figure}

{The left plot in Figure \ref{fig:power-vs-temperature} shows the outdoor unit's hourly average input power vs. the hourly average indoor-outdoor temperature difference, calculated using Equation \eqref{eq:HDH}, for AC (blue) and DC (red) testing. The indoor-outdoor temperature difference is a reasonable stand-in for heat demand, which for most buildings scales approximately linearly with the temperature difference between the indoor and outdoor air.} The left plot in Figure \ref{fig:power-vs-temperature} also shows quadratic fits and 90\% Gaussian confidence intervals. The heat pump appears to draw slightly less power on DC than on AC for both low and high temperature differences. For intermediate temperature differences, the heat pump appears to draw slightly more power on AC than on DC.

The right plot in Figure \ref{fig:power-vs-temperature} shows outdoor unit's daily average input power vs. the daily average indoor-outdoor temperature difference for AC and DC. Figure \ref{fig:power-vs-temperature} also shows linear fits (dashed lines) and 90\% Gaussian confidence intervals (shaded areas). {The $R^2$ values for the AC and DC fits are 0.895 and 0.931, respectively. The AC fit likely has a higher $R^2$ value because there are more AC data at low indoor-outdoor temperature differences. This is because the weather was somewhat warmer during AC testing than DC testing. DC testing included more very cold days, and the associated data points tend to increase the slope of the DC best-fit line.} Although there were slight performance differences on AC and DC power in field, the overlapping confidence intervals suggest that the performance differences are difficult to distinguish from the measurement uncertainty ranges.

{ A statistical hypothesis test was conducted to formally evaluate whether the field data indicate that the heat pump performed differently on AC vs. DC. The hypothesis test involved the weather-normalized daily average input powers on AC and DC,
\begin{equation}
X_a = \frac{ P_a }{ \Delta T - c } , \ X_d = \frac{ P_d }{ \Delta T - c } ,
\end{equation}
where $P_a$ (kW) and $P_d$ (kW) are the daily average input powers on AC and DC, respectively, $\Delta T$ ($^\circ$C) is the daily average indoor-outdoor temperature difference, and $c =$ 8 $^\circ$C is the indoor-outdoor temperature difference at which heat demand is approximately zero. The random variables $X_a$ and $X_d$ were modeled as independent and Gaussian with potentially different means and variances. Each $(\Delta T_a, P_a)$ observation on AC gave rise to a realization of the corresponding random variable $X_a$, and similarly for DC. The null and alternative hypotheses were
\begin{equation}
H_0: \mu_a = \mu_d, \ H_1: \mu_a \neq \mu_d ,
\end{equation}
where $\mu_a$ and $\mu_d$ are the population means of $X_a$ and $X_d$, respectively. Under this model, the hypothesis test was a two-sample two-tailed Welch's unequal variance $t$-test with test statistic
\begin{equation}
t = \frac{ \bar X_a - \bar X_d}{ \sqrt{s_a^2 / N_a + s_d^2 / N_d} } .
\end{equation}
Here $\bar X_a$ and $\bar X_d$ are the sample means, $s_a^2$ and $s_d^2$ are the unbiased sample variances, and $N_a$ and $N_d$ are the sample sizes. }

{The hypothesis test results were computed in Python using the SciPy package's {\it ttest\_ind} function with the {\it equal\_var = False} option. The resulting $p$-value, meaning the probability that the difference in sample means is greater than or equal to the observed difference, assuming the population means are equal, was 0.18. Because $0.23 > 0.05$, the hypothesis test indicates that the difference in weather-normalized daily-average input powers on AC vs. DC was not statistically significant at a significance level of $\alpha = 0.05$.}

\section{System-level nanogrid modeling}
\label{sys-mod}

This section presents the general nanogrid model equations and control logic (Section \ref{generalModel}), the simulated AC and DC nanogrid configurations (Section \ref{nanogridConfigurations}), and the simulation results (Section \ref{simulationResults}).

\subsection{General nanogrid model}
\label{generalModel}

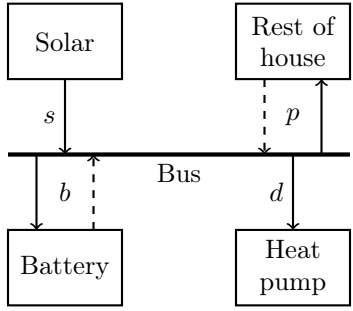
\begin{figure}
\centering
\begin{tikzpicture}[xscale=1.5]
    \draw[ultra thick] (0,0) -- (3,0);
    \node[below] at (1.5,0) {Bus};
    
    \draw[thick] (0,1) rectangle (1,2);
    \node at (0.5,1.5) {Solar};
    \draw[thick, ->] (0.5,1) -- (0.5,0);
    \node[left] at (0.5,0.5) {$s$};

    \draw[thick] (0,-1) rectangle (1,-2);
    \node at (0.5,-1.5) {Battery};
    \draw[thick, ->] (0.25,0) -- (0.25,-1);
    \node at (0.5,-0.5) {$b$};
    \draw[thick, dashed, ->] (0.75,-1) -- (0.75,0);

    \draw[thick] (2,1) rectangle (3,2);
    \node[align=center] at (2.5,1.5) {Rest of \\ house};
    \draw[thick, ->] (2.75,0) -- (2.75,1);
    \node at (2.5,0.5) {$p$};
    \draw[thick, dashed, ->] (2.25,1) -- (2.25,0);

    \draw[thick] (2,-1) rectangle (3,-2);
    \node[align=center] at (2.5,-1.5) {Heat \\ pump};
    \draw[thick, ->] (2.5,0) -- (2.5,-1);
    \node[left] at (2.5,-0.5) {$d$};

\end{tikzpicture}
\caption{Power flows in the nanogrid include the solar output $s$, battery charging (or if negative, discharging) $b$, and heat pump input $d$. From a power balance on the bus, the nanogrid exports (or if negative, imports) $p = s - d - b$ to (or from) the rest of the house.}
\label{nanogridArchitecture}
\end{figure}

Figure \ref{nanogridArchitecture} shows the general nanogrid architecture, which includes solar PV, a stationary battery with energy and power capacities $\overline x = 20$ kWh and $\overline b = 12.5$ kW, respectively, a heat pump, and a bidirectional connection to distribution system that serves the rest of the house. These components are all connected on a common bus. The solar power output $s > 0$ (kW) is injected to the bus. The battery's electric charging (or if negative, discharging) power $b$ (kW) is extracted from (or injected to) the bus. The heat pump's electric power $d > 0$ (kW) is extracted from the bus. From a power balance on the bus, the power that the nanogrid exports to (or if negative, imports from) the rest of the house is $p = s - b - d$ (kW).

The battery is modeled as a first-order linear dynamical system governed by the ordinary differential equation
\begin{equation}\label{eq:battery-ode}
    \frac{\text d x(t)}{\text d t} = - \frac{x(t)}{\tau} + u(t) ,
\end{equation}
where $t$ (h) denotes time, $x$ (kWh) is the chemical energy stored in the battery, the self-dissipation time constant is $\tau = 1600$ h (a reasonable value for lithium-ion batteries), and $u$ (kW) is the chemical charging (or if negative, discharging) power. Assuming the charging power is piecewise constant over each discrete time step of duration $\Delta t = 1$ h, the battery dynamics \eqref{eq:battery-ode} discretize exactly to
\begin{equation}\label{eq:battery-model}
    x(k+1) = a x(k) + (1 - a)\tau u(k) ,
\end{equation}
where the integer $k$ indexes discrete time steps and $a = \exp(-\Delta t / \tau)$. The electrical charging power $b$ (kW) is modeled as a piecewise linear function of the chemical charging power:
\begin{equation}\label{eq:battery-power}
\begin{aligned}
    b(k) &= 
    \begin{cases}
            \eta u(k) &\quad \text{if } u(k) \leq 0 \\
            u(k)/ \eta  &\quad \text{if } u(k) > 0  \\
    \end{cases} \\
    &= \max(\eta u(k), u(k) / \eta) ,
\end{aligned}
\end{equation}
where $\eta = 0.95$ is the battery's charging and discharging efficiency. Under this model, injecting a chemical charging power of 1 kW to the battery requires extracting $1/\eta$ kW of electric power from the bus. Similarly, discharging 1 kW of chemical power from the battery injects $\eta$ kW of electric power to the bus.

The control logic prioritizes using solar power first for the heat pump, then to charge the battery (if the solar supply $s$ exceeds heat pump demand $d$), then for exports to the rest of the house. If the heat pump demand exceeds the solar supply, the control logic prioritizes powering the heat pump first from the battery, then (if the battery is empty) by importing power from the rest of the house. At each time step, the control logic also ensures that the energy stored at the next time step, $x(k+1)$ in Equation \eqref{eq:battery-model}, remains above zero and below the energy capacity $\overline x$. Similarly, the control logic ensures that the electric charging power, $b(k)$ from Equation \eqref{eq:battery-power}, remains above $-\overline b$ and below $\overline b$. The control logic can be written concisely as
{ \begin{equation}
    u(k) = \begin{cases}
        \eta \min\left( s(k) - d(k), \overline b, (\overline x - a x(k))/((1-a)\tau) \right) \quad &\text{if } s(k) \geq d(k) \\
        (1/\eta) \max\left( s(k) - d(k), -\overline b, -a x(k)/((1-a)\tau) \right) &\text{otherwise.} \\
    \end{cases}
\end{equation} }

The input signals to the model are the solar power output $s$ and the heat pump demand $d$. The solar array is modeled on the rooftop PV system at the DC Nanogrid House, which consists of four sub-arrays with the parameters in Table \ref{ta:solar-array}. The array has a total nameplate capacity of 14.3 kW DC. The solar power output is simulated using PVLib, an open source Python library \cite{anderson_pvlib_2023}, and solar irradiance data from the Oikolab weather service. The simulated heat pump is the same unit from the field testing in this paper. It has 14 kW (4 tons) of nameplate cooling capacity. As the laboratory and field test results suggest that the heat pump performance between AC and DC is equivalent within 5 to 10\%, which is comparable to the measurement uncertainty range, the simulations use the same historical heat pump power data to represent its electricity demand $d$ on either AC or DC. {The simulations also use historical measurements of the rest-of-house AC load, including the auxiliary heating elements. Monthly electric bills are calculated and compared across nanogrid configurations. When solar supply exceeds whole-house demand, the utility is assumed to buy the surplus at the full retail price of 0.14 USD/kWh \cite{duke-price}.}

\begin{table}
  \caption{Solar array characteristics from the DC Nanogrid House}
  \vspace*{6pt}
  \centering
    \begin{tabular}{@{}cccc@{}}
    \textbf{Sub-array} & \textbf{Tilt [deg]} & \textbf{Azimuth [deg]} & \textbf{Modules [\#]} \\
    \hline
    1 & 32 & 90 & 3 \\
    2 & 50 & 180 & 3 \\
    3 & 32 & 90 & 6 \\
    4 & 30 & 270 & 30 \\
    \end{tabular}
  \label{ta:solar-array}
\end{table}

\subsection{AC and DC nanogrid configurations}
\label{nanogridConfigurations}

\begin{figure}
  \centering
  \includegraphics[width=0.9\columnwidth]{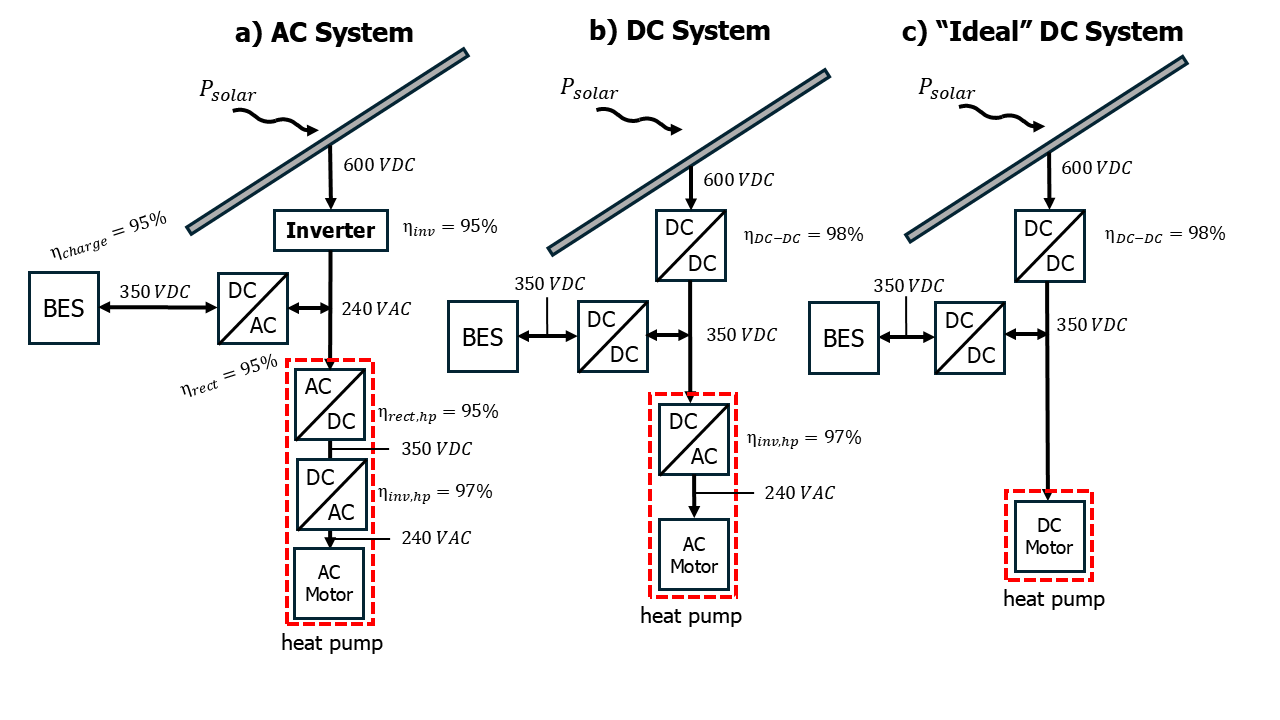}
  \caption{AC and DC configurations for connecting the PV, battery, and heat pump in a nanogrid.}
  \label{fig:nanogrid-configs}
\end{figure}

\begin{figure}
  \centering
  \includegraphics[width=0.7\columnwidth]{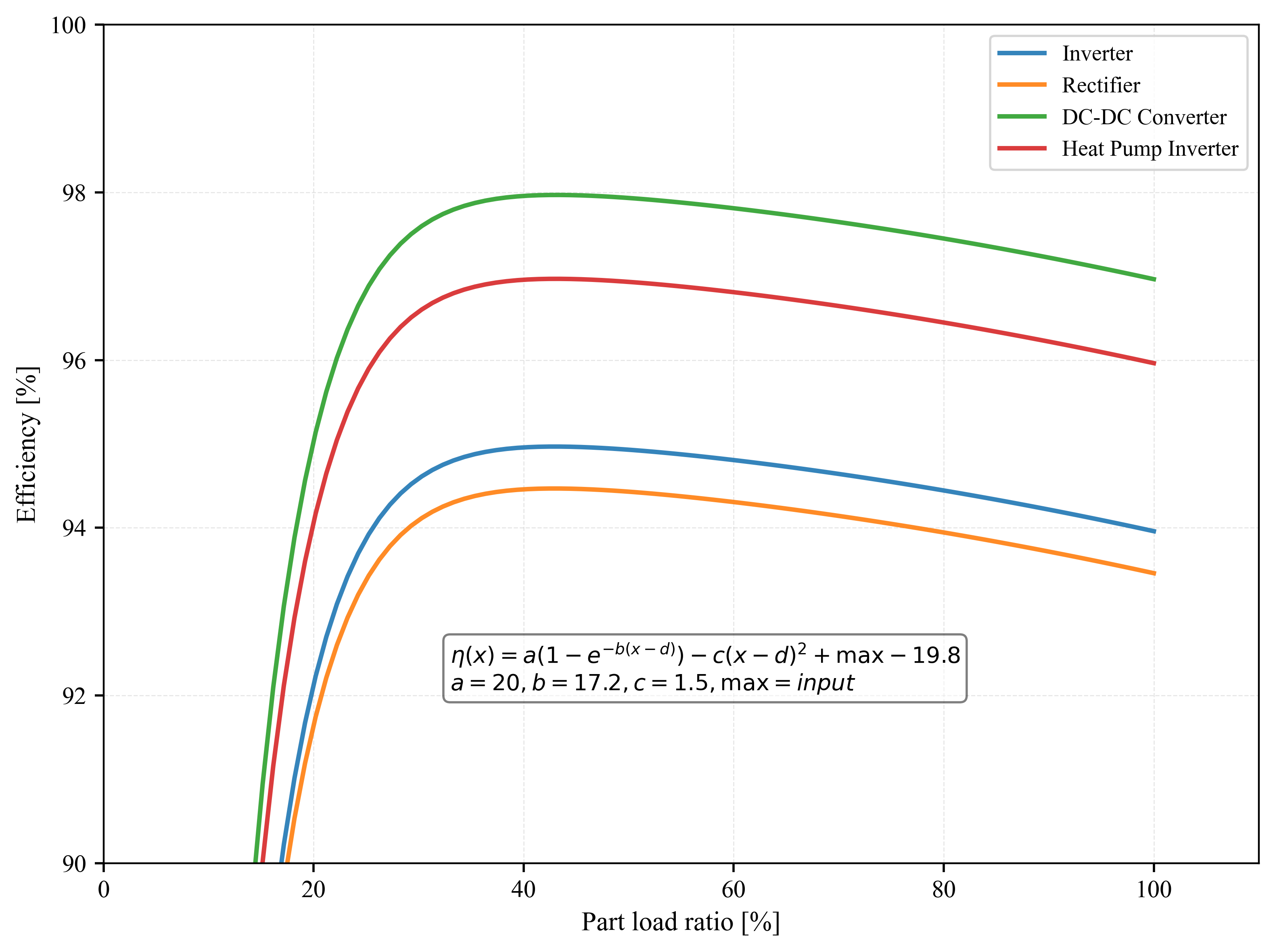}
  \caption{Part-load efficiency curves for each converter in the modeled nanogrid configurations.}
  \label{fig:efficiencies}
\end{figure}

Figure \ref{fig:nanogrid-configs} shows the three nanogrid configurations modeled here. The first (left) configuration is a typical AC nanogrid that includes PV and battery, a conventional AC bus, and a heat pump. The battery and PV interface with the AC bus through inverters. The heat pump internally rectifies the 60 Hz AC from the bus, then inverts the DC to variable-frequency AC so that the AC motor can drive the compressor at a desired speed. This heat pump configuration represents the hardware that was field-tested at the DC Nanogrid House. The second (center) configuration is similar to the first, but the PV, battery, and heat pump connect on a 350 V DC bus, each through their own DC-DC converter. The PV DC-DC converter implements maximum power-point tracking, stepping down the peak 600 V DC {\color{black} PV input voltage} to 350 V DC. The battery converter implements charge control. At the heat pump, the DC power bypasses the rectifier and feeds into the same internal inverter depicted in the left configuration. The third (right) configuration is similar to the second, but the heat pump's AC motor is replaced by a DC motor that connects directly to the DC bus. The nanogrid configurations in Figure \ref{fig:nanogrid-configs} are based in part on \cite{vossos-2017-DC-appliances}. Converter efficiencies are modeled using the part-load curves illustrated in Figure \ref{fig:efficiencies}. The peak of each converter's part-load curve was tuned to a value obtained from a manufacturer specification sheet for a representative converter of the same type and cross-checked against the converters used in the real nanogrid. Table \ref{ta:conversion-eff} shows the peak efficiency for each converter.

\begin{table}
  \caption{Max converter efficiencies (from Vossos et al.  \cite{vossos-2017-DC-appliances})}
  \vspace*{6pt}
  \centering
    \begin{tabular}{@{}l l l@{}}
    \textbf{Component} & \textbf{Peak efficiency [\%]} \\
    \midrule
    Inverter & 95 \\
    Rectifier & 95 \\
    Maximum power-point tracker & 98 \\
    DC-DC converter & 98 \\
    Heat pump inverter & 97 \\
    \end{tabular}
  \label{ta:conversion-eff}
\end{table}

\subsection{Simulation results}
\label{simulationResults}

\begin{figure}
  \centering
  \includegraphics[width=0.8\columnwidth]{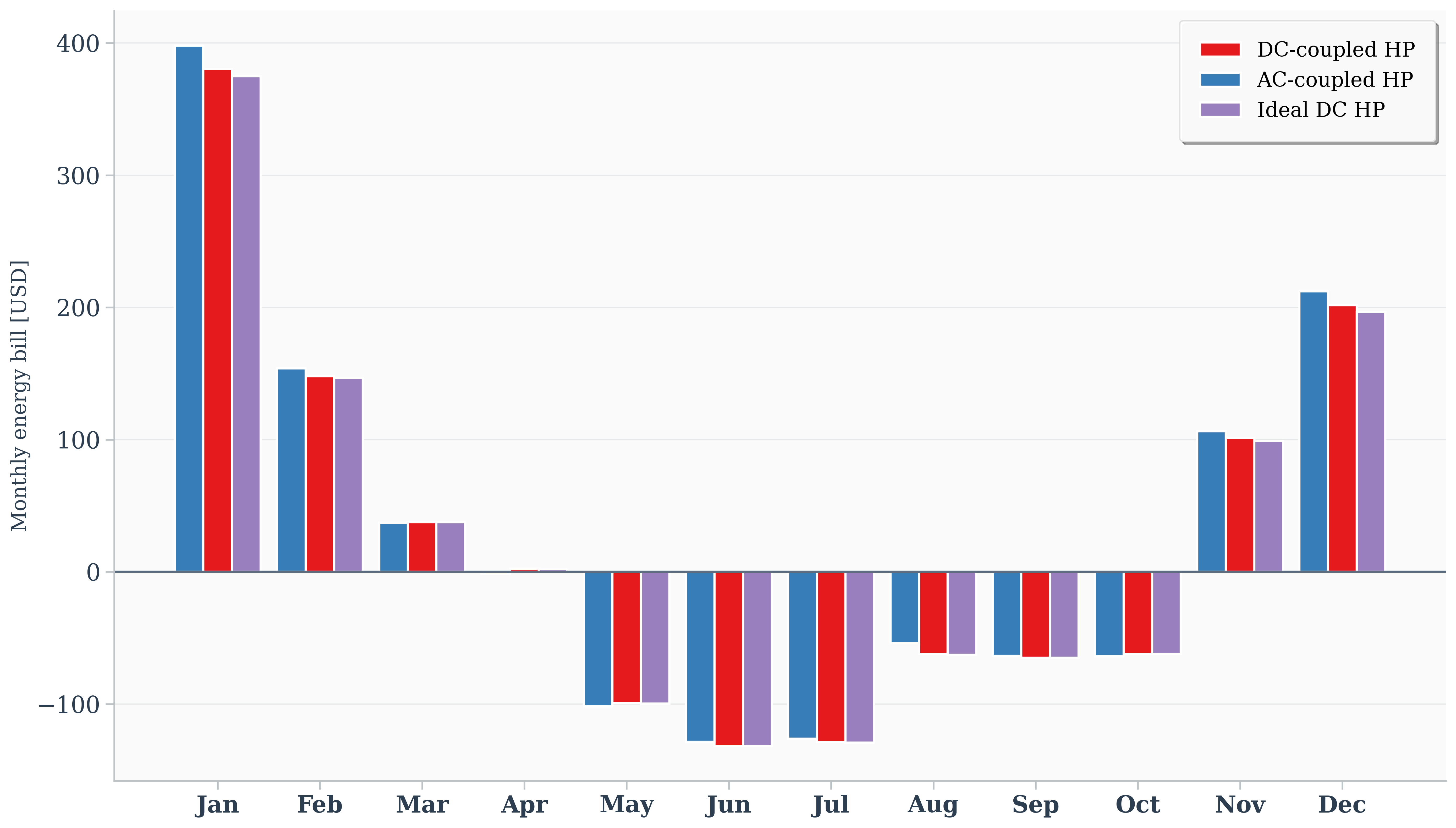}
  \caption{Monthly {electricity bills for the test house} under the three nanogrid configurations in Figure \ref{fig:nanogrid-configs}. {Electricity bills are highest in winter, when heat demand is highest and PV supply is lowest.}}
  \label{fig:monthly-cost}
\end{figure}

The three nanogrid configurations discussed in Section \ref{nanogridConfigurations} were simulated using the equations and logic from Section \ref{nanogridModeling} under coincident hourly data from the calendar year of 2024. {Figure \ref{fig:monthly-cost} shows the monthly electric bills based on simulations that include measured historical power data for the heat pump and the rest-of-house loads, as well as modeled power flows in the nanogrid. Monthly bills are highest in January and February, when heat demand is highest and PV supply is lowest. Winter bills are highest for the AC nanogrid configuration, which has the highest converter losses, and are lowest for the ideal DC configuration. In the summer, the house exports surplus solar power, leading to negative bills (payments from the utility to the house's residents). The payments from the utility are lowest in the AC nanogrid configuration and highest in the ideal DC configuration.} 

\begin{figure}
  \centering
  \includegraphics[width=0.65\columnwidth]{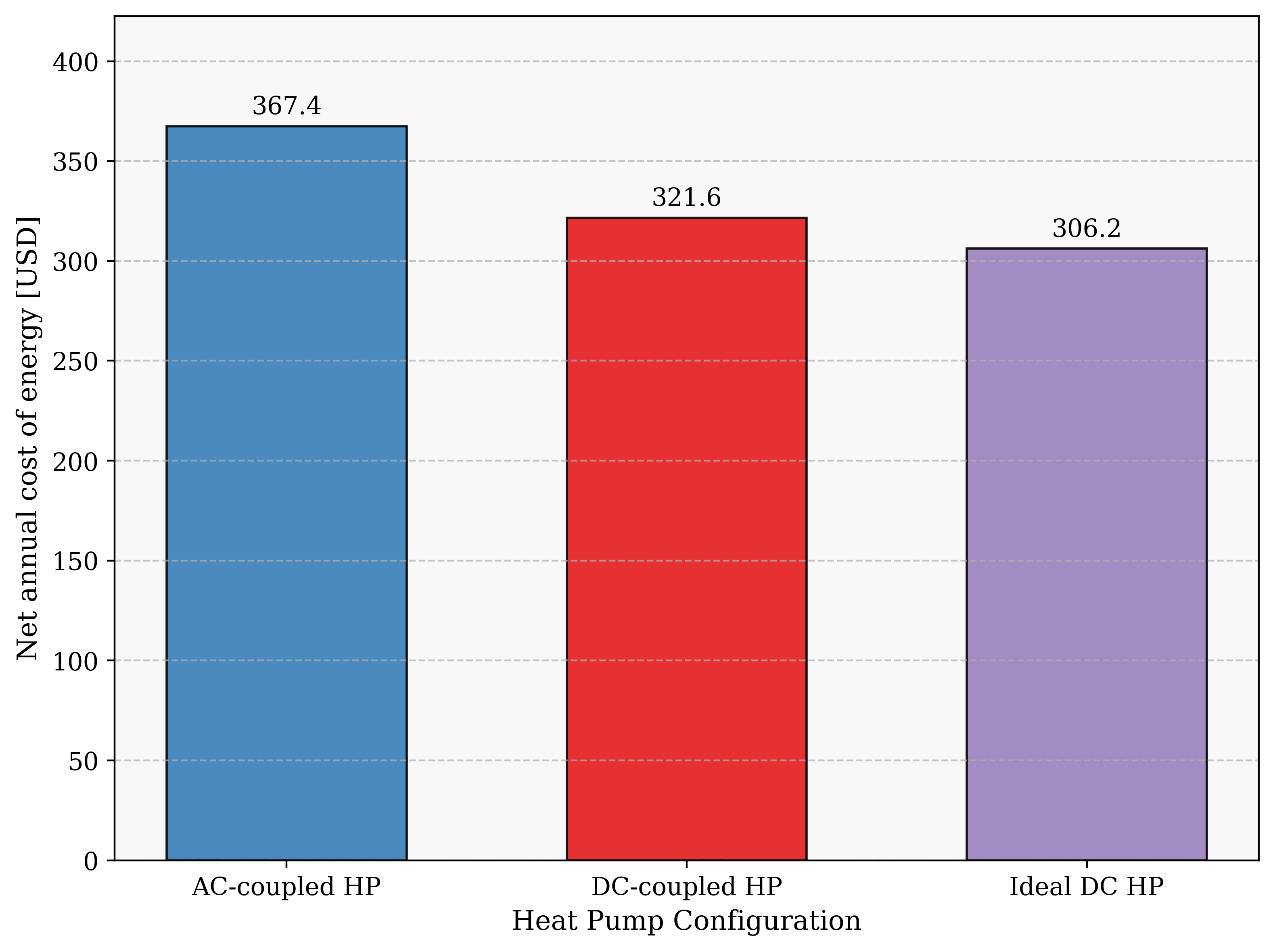}
  \caption{Net annual {electricity bill for} the test house under the three nanogrid configurations in Figure \ref{fig:nanogrid-configs}. Annual bills are highest in the AC configuration and lowest in the ideal DC configuration.}
  \label{fig:year-cost}
\end{figure}

{Figure \ref{fig:year-cost} shows the annual electricity bill estimates for the three configurations. The annual bill estimate is highest in the AC configuration (left, 367.4 USD), which has the highest conversion losses. The annual bill estimate for the after-market DC heat pump retrofit that was actually implemented in the test house (center, 321.6 USD) was 12.5\% lower than the AC bill estimate. In the ideal DC configuration (right, 306.2 USD), the annual bill estimate was 16.7\% lower than the AC bill estimate.}

The laboratory and field testing results from Section \ref{exp-results} informed the nanogrid modeling in this section. Although device-level testing indicated no statistically significant differences in operating the heat pump on AC vs. DC, modeling indicated significant energy and cost savings from converting the nanogrid as a whole to DC. Most of these benefits come from the shifting from the first configuration in Figure \ref{fig:nanogrid-configs} to the second, which integrates PV and battery power on a common DC bus but runs the heat pump internally on AC,  as shown in Figure \ref{fig:year-cost}. The incremental improvement from moving from the second configuration to the third, which runs the heat pump internally on DC, are small by comparison. 

\section{Discussion}
\label{discussion}
This section summarizes this paper's main results (Section \ref{synthesisResults}), discusses limitations of the experimental and modeling work (Section \ref{limitations}), and highlights areas of future work (Section \ref{futureWork}).

\subsection{Synthesis of results}
\label{synthesisResults}

This study presented three comparisons of AC and DC operation of an off-the-shelf residential unitary air-source heat pump. The first comparison used steady-state laboratory testing under the AHRI Standard 210/240 conditions. The second comparison used field testing in an occupied house with another heat pump of the same make and model as the one tested in the laboratory. The third comparison used modeling of a residential nanogrid with solar PV and a stationary battery. The laboratory tests showed that under steady-state operation, the heat pump performed equivalently under AC and DC within 6\%, a range comparable with measurement uncertainty. The COP appeared to be marginally higher on AC at low load, but marginally higher on DC at high load. However, as noted in the sensor accuracy analysis, relative measurement uncertainties were larger at low load. The field tests broadly corroborated the laboratory findings. When comparing heat pump operation on AC vs. DC under similar boundary conditions for one month, the weather-normalized input power was approximately the same to within 10\%. This range is again comparable to the measurement uncertainty range, which is somewhat higher for field testing than for laboratory testing.

Although switching the heat pump from AC to DC did not affect performance significantly at the device level, the switch did show efficiency improvements in system-level nanogrid modeling. In simulations, the DC nanogrid configuration reduced net annual electricity bills by 12.5\% to 16.7\%. The lower electricity bills in the DC configurations came from decreasing the number of power conversions and from increasing converter efficiencies (and therefore the battery's round-trip efficiency). The 12.5\% to 16.7\% savings refer to the net (demand minus supply) energy used by the building. The gross (demand only) energy used by the nanogrid components -- including the battery, heat pump, and converters -- was 9.1 MWh for the AC configuration, 8.4 MWh for the intermediate DC configuration, and 8.3 MWh for the ideal DC configuration. The gross energy savings from converting the DC nanogrid from AC to DC were therefore 8\% and 9.2\%. For comparison, \cite{alshammari2021dcIreland} reported 5\% savings from a modeled DC project in Ireland.

For the 208 m$^2$, all-electric, detached single-family house in Indiana considered in this study, the estimated annual electricity bill savings from an ideal DC nanogrid configuration, relative to a standard AC configuration, were 61.2 USD. With these annual bill savings, and assuming a simple payback of no more than ten years is required, the highest justifiable capital cost of retrofitting a residential nanogrid to run on DC is 612 USD. Today's DC retrofit costs are likely far higher than this ceiling, and therefore not justifiable based on savings alone. However, DC retrofit costs may come down in the future. DC may also make financial sense in new residential construction, where the costs of installing a new DC distribution system are likely comparable to those of installing a new AC distribution system, or for retrofits in larger buildings -- such as commercial, industrial, or multifamily residential -- where the modest percent savings estimated in this paper translate to larger absolute savings. Increased adoption of electric vehicles could significantly increase household energy demand and improve the economic case for DC distribution.

\subsection{Limitations}
\label{limitations}

This paper has several limitations. As previously mentioned, the heat pump compressor drive was not modified to natively accept DC input in the laboratory or field testing. At the manufacturer's recommendation, DC power was simply fed into the same terminals into which AC power would normally connect. On DC, the heat pump internally bypassed its rectifier, feeding the bus's DC power directly to the heat pump's on-board inverter. {This could be one source of the differences in the laboratory and field studies, which show the heat pump using slightly more power on DC than on AC in some scenarios. For instance, under DC operation, the bridge-rectifier may have dropped the DC voltage going into the variable-speed drive, decreasing efficiency and potentially accelerating equipment degradation.} More detailed testing should be conducted with electrical measurements along the compressor drive to understand the losses which occur between AC and DC power inputs.

A persistent challenge in the field study was normalizing heat pump's performance for differences in weather conditions, occupancy patterns, and possibly equipment degradation between the AC and DC testing. This paper normalized performance data by indoor-outdoor temperature differences, but future work could investigate other normalization methods. 

{Better modeling of converter part-load efficiencies could refine the system-level nanogrid modeling results. This study used a single, stylized part-load curve for all converters, shifted vertically to model different peak efficiencies for different converters. A more sophisticated approach would tailor each converter's full part-load efficiency curve to data from manufacturer specification sheets or from field measurements. This study also neglected real-world effects in DC microgrids and nanogrids, such as voltage fluctuations, harmonic effects, load-varying stability, grid interactions, and fault protection. Future work could investigate these effects.}

\subsection{Future work}
\label{futureWork}

{Future work could build on this study in at least three ways. First, a nanogrid with, at minimum, a heat pump, a stationary battery, and a PV solar array could be tested in the field. This would allow refined modeling of nanogrid components, experimental investigation of effects that were not modeled here (such as stability, harmonics, and protection), and real-world observation of annual electricity bills. The nanogrid could also be expanded to include an electric vehicle charger or other large loads. Second, after-market DC heat pump conversions could be investigated in more detail. For example, the retrofitted heat pump's variable-speed drive and compressor could be placed on a hot-gas bypass test stand to more thoroughly and precisely evaluate performance. Future work could also test heat pumps with natively DC motors (fixed-speed, two-stage, or variable-speed), rather than the heat pump tested here, which in its default configuration rectified the input AC power then inverted the resulting DC power to a desired frequency in order to govern the compressor speed. Third, field data on DC nanogrid deployment costs could be gathered, enabling cost-benefit analysis and highlighting important areas for innovation.}

\section{Conclusion}
\label{conclusion}

A growing body of simulation research suggests that connecting natively DC devices -- such as solar photovoltaics, electric vehicle chargers, stationary batteries, and major appliances driven by DC motors -- through DC distribution within buildings could improve energy efficiency. However, this hypothesis remains largely untested in hardware. Toward addressing that gap, this paper tested a major appliance -- an air-source heat pump used for space heating and cooling -- on AC and DC in the laboratory and the field. {As far as the authors are aware, this study reported the first laboratory testing of a DC heat pump retrofit in heating mode, and the first DC heat pump retrofit field testing of any kind. These first-of-a-kind experiments suggested that retrofitting the heat pump from AC to DC did not significantly alter energy efficiency at the device level. However, calibrated year-long system-level modeling suggested that connecting the heat pump to solar photovoltaics and a battery in a DC nanogrid configuration could decrease the test house's annual electricity bills by 12.5\% to 16.7\%. To the authors' knowledge, these are the first modeling results informed by real-world data from an occupied test house, and therefore constitute a meaningful step toward trustworthy quantification of the benefits of DC distribution systems in residential buildings.}

\printnomenclature

\section*{Declaration of Competing Interest}
The authors declare that they have no competing financial interests or personal relationships that could have appeared to influence the work reported in this paper.

\section*{Declaration of Generative AI and AI-assisted technologies in the writing process}
Aaron H.P. Farha used Claude Sonnet 4.0 to format matplotlib plots in Python. Aaron H.P. Farha reviewed and edited the Claude output and takes full responsibility for the content of the publication.

\section*{CRediT Author Statement}
\textbf{Aaron H.P. Farha:} Methodology, Investigation, Software, Visualization, Conceptualization, Writing - Original Draft, Writing - Reviewing and Editing. \textbf{Jonathan P. Ore:} Writing - Reviewing and Editing, Data Curation. \textbf{Elias N. Pergantis:} Methodology, Data Curation, Conceptualization, Reviewing and Editing. \textbf{Davide Ziviani:} Writing - Reviewing and Editing, Conceptualization, Project Administration. \textbf{Eckhard A. Groll:} Writing - Reviewing and Editing, Project Administration, Funding Acquisition. \textbf{Kevin J. Kircher:} Writing - Reviewing and Editing, Visualization, Project Administration, Conceptualization.

\section*{Data availability}

Data and code are publicly available at \url{https://github.com/AaronFarha/dchp}.

\section*{Acknowledgments}
The Center for High Performance Buildings at Purdue University supported this work. The authors thank Trane Technologies for technical support during laboratory testing and field retrofits, the DC Nanogrid House residents for their patience during field testing, and  Alex Lee for code review. {\color{black} We thank the anonymous peer reviewers for providing detailed and thorough feedback that substantially improved the work.}


\printbibliography

@misc{ahri-210-240,
   author = {{American Heating, Air-Conditioning and Refrigeration Institute}},
   title = {Performance Rating of Unitary Air-conditioning \& Air-source Heat Pump Equipment},
   howpublished = {AHRI Standard 210/240},
   year = {2023},
   organization = {American Heating, Air-Conditioning and Refrigeration Institute}
}

@article{assaf_power_2024,
	title = {Power Converter Topologies for Heat Pumps Powered by Renewable Energy Sources: A Literature Review},
	volume = {13},
	rights = {https://creativecommons.org/licenses/by/4.0/},
	issn = {2079-9292},
	url = {https://www.mdpi.com/2079-9292/13/19/3965},
	doi = {10.3390/electronics13193965},
	pages = {3965},
	number = {19},
	journaltitle = {Electronics},
	author = {Assaf, Joyce and Menye, Joselyn Stephane and Camara, Mamadou Baïlo and Guilbert, Damien and Dakyo, Brayima},
	urldate = {2025-07-16},
	date = {2024-10-09},
	langid = {english},
	note = {Publisher: {MDPI} {AG}}
}

@inbook{chauhan_dc_2018,
	title = {{DC} microgrid in residential buildings},
	url = {https://digital-library.theiet.org/content/books/10.1049/pbpo115e\_ch15},
	pages = {367--388},
	booktitle = {{DC} Distribution Systems and Microgrids},
	publisher = {Institution of Engineering and Technology},
	author = {Chauhan, Rajeev Kumar and Gonzalez-Gongatt, Francisco and Rajpurohit, Bharat Singh and Singh, Sri Nivas},
	urldate = {2025-07-16},
	date = {2018-10-05},
	langid = {english},
	doi = {10.1049/pbpo115e\_ch15}
}

@article{chen_networked_2021,
	title = {Networked Microgrids for Grid Resilience, Robustness, and Efficiency: A Review},
	volume = {12},
	rights = {https://ieeexplore.ieee.org/Xplorehelp/downloads/license-information/{IEEE}.html},
	issn = {1949-3053, 1949-3061},
	url = {https://ieeexplore.ieee.org/document/9170837/},
	doi = {10.1109/tsg.2020.3010570},
	pages = {18--32},
	number = {1},
	journaltitle = {{IEEE} Trans. Smart Grid},
	author = {Chen, Bo and Wang, Jianhui and Gu, Xiaonan and Chen, Chen and Zhao, Shijia},
	urldate = {2025-07-16},
	date = {2021-01},
	langid = {english},
	note = {Publisher: Institute of Electrical and Electronics Engineers ({IEEE})}
}

@article{dragicevic_dc_2016,
	title = {{DC} Microgrids—Part {II}: A Review of Power Architectures, Applications, and Standardization Issues},
	volume = {31},
	rights = {https://ieeexplore.ieee.org/Xplorehelp/downloads/license-information/{IEEE}.html},
	issn = {0885-8993, 1941-0107},
	url = {http://ieeexplore.ieee.org/document/7177102/},
	doi = {10.1109/tpel.2015.2464277},
	pages = {3528--3549},
	number = {5},
	journaltitle = {{IEEE} Trans. Power Electron.},
	author = {Dragicevic, Tomislav and Gu, Xiaonan and Vasquez, Juan C. and Guerrero, Josep M.},
	urldate = {2025-07-16},
	date = {2016-05},
	langid = {english},
	note = {Publisher: Institute of Electrical and Electronics Engineers ({IEEE})},
}

@misc{eia-2025-energy-use,
    author = {{U.S. Energy Information Administration}},
    title = {Energy Consumption by Sector},
    year = {2025},
    url = {http://www.eia.gov/totalenergy/data/monthly/\#consumption}
}

@article{CHARALAMBOUS-2023,
title = {Hybrid {AC}-{DC} distribution system for building integrated photovoltaics and energy storage solutions for heating-cooling purposes. A case study of a historic building in {C}yprus},
journaltitle = {Renewable Energy},
volume = {216},
pages = {119032},
year = {2023},
issn = {0960-1481},
doi = {https://doi.org/10.1016/j.renene.2023.119032},
url = {https://www.sciencedirect.com/science/article/pii/S0960148123009461},
author = {Chrysanthos Charalambous and Chryso Heracleous and Aimilios Michael and Venizelos Efthymiou}
}

@article{elsayed_dc_2015,
	title = {{DC} microgrids and distribution systems: An overview},
	volume = {119},
	rights = {https://www.elsevier.com/tdm/userlicense/1.0/},
	issn = {0378-7796},
	url = {https://linkinghub.elsevier.com/retrieve/pii/S0378779614003885},
	doi = {10.1016/j.epsr.2014.10.017},
	pages = {407--417},
	journaltitle = {Electric Power Systems Research},
	author = {Elsayed, Ahmed T. and Mohamed, Ahmed A. and Mohammed, Osama A.},
	urldate = {2025-07-16},
	date = {2015-02},
	langid = {english},
	note = {Publisher: Elsevier {BV}},
}

@report{garbesi_optimizing_nodate,
	title = {Optimizing Energy Savings from Direct-{DC} in {U.S.} Residential Buildings},
	author = {Garbesi, Karina and Vossos, Vagelis and Sanstad, Alan and Burch, Gabriel},
	langid = {english},
    date={2012-07-16},
	institution = {Lawrence Berkeley National Laboratory ({LBNL})},
}

@report{garbesi_catalog_2010,
	location = {Berkeley, {CA} (United States)},
	title = {Catalog of {DC} Appliances and Power Systems},
	url = {https://www.osti.gov/servlets/purl/1076790/},
	institution = {Lawrence Berkeley National Laboratory ({LBNL})},
	author = {Garbesi, Karina and Vossos, Vagelis and Shen, Hongxia},
	urldate = {2025-07-16},
	date = {2010-10-13},
	langid = {english},
	doi = {10.2172/1076790},
}

@inproceedings{kakigano_loss_2010,
	location = {Sapporo, Japan},
	title = {Loss evaluation of {DC} distribution for residential houses compared with {AC} system},
	url = {http://ieeexplore.ieee.org/document/5543501/},
	doi = {10.1109/ipec.2010.5543501},
	eventtitle = {2010 International Power Electronics Conference ({IPEC} - Sapporo)},
	booktitle = {The 2010 International Power Electronics Conference - {ECCE} {ASIA} -},
	publisher = {{IEEE}},
	author = {Kakigano, H. and Nomura, M. and Ise, T.},
	urldate = {2025-07-16},
	date = {2010-06},
	langid = {english},
}

@article{shahidehpour_microgrids_2016,
	title = {Microgrids for Enhancing the Power Grid Resilience in Extreme Conditions},
	rights = {https://ieeexplore.ieee.org/Xplorehelp/downloads/license-information/{IEEE}.html},
	issn = {1949-3053, 1949-3061},
	url = {http://ieeexplore.ieee.org/document/7489002/},
	doi = {10.1109/tsg.2016.2579999},
	pages = {1--1},
	journaltitle = {{IEEE} Trans. Smart Grid},
	author = {Shahidehpour, Mohammad and Liu, Xindong and Li, Zuyi and Cao, Yijia},
	urldate = {2025-07-16},
	date = {2016},
	langid = {english},
	note = {Publisher: Institute of Electrical and Electronics Engineers ({IEEE})},
}

@article{muruganantham_challenges_2017,
	title = {Challenges with renewable energy sources and storage in practical distribution systems},
	volume = {73},
	rights = {https://www.elsevier.com/tdm/userlicense/1.0/},
	issn = {1364-0321},
	url = {https://linkinghub.elsevier.com/retrieve/pii/S1364032117301004},
	doi = {10.1016/j.rser.2017.01.089},
	pages = {125--134},
	journaltitle = {Renewable and Sustainable Energy Reviews},
	author = {Muruganantham, B. and Gnanadass, R. and Padhy, N.P.},
	urldate = {2025-07-16},
	date = {2017-06},
	langid = {english},
	note = {Publisher: Elsevier {BV}},
}

@phdthesis{ore-thesis-2021,
    author = {Jonathan Paul Ore},
    title = {The {DC} {N}anogrid {H}ouse: Converting a Residential Building from {AC} to {DC} Power to Improve Energy Efficiency},
    school = {Purdue University},
    year = {2021}
}

@article{ore_evaluation_2020,
	title = {Evaluation of a Hybrid {AC}/{DC} Powered Residential Split- System Heat Pump Performance using a {DC} Nanogrid},
	author = {Ore, Jonathan and Meral, Fatih and Obst, Oliver and Kurtulus, Orkan and Groll, Eckhard A},
	date = {2020},
	langid = {english},
    journaltitle = {{International} {Energy} {Agency}}
}

@article{stippich_ac_2017,
	title = {From {AC} to {DC}: Demonstration of Beneﬁts in Household Appliances},
	author = {Stippich, Alexander and Sewergin, Alexander and Engelmann, Georges and Gottschlich, Jan and Neubert, Markus},
	date = {2017},
	langid = {english},
}

@article{vossos_energy_2014,
	title = {Energy savings from direct-{DC} in {U.S.} residential buildings},
	volume = {68},
	rights = {https://www.elsevier.com/tdm/userlicense/1.0/},
	issn = {0378-7788},
	url = {https://linkinghub.elsevier.com/retrieve/pii/S0378778813005720},
	doi = {10.1016/j.enbuild.2013.09.009},
	pages = {223--231},
	journaltitle = {Energy and Buildings},
	author = {Vossos, Vagelis and Garbesi, Karina and Shen, Hongxia},
	urldate = {2025-07-16},
	date = {2014-01},
	langid = {english},
	note = {Publisher: Elsevier {BV}},
}

@inproceedings{ore_analysis_2020,
	location = {The Hague, Netherlands},
	title = {Analysis of a Residential House for the Design and Implementation of a {DC} Nanogrid},
	rights = {https://ieeexplore.ieee.org/Xplorehelp/downloads/license-information/{IEEE}.html},
	url = {https://ieeexplore.ieee.org/document/9248788/},
	doi = {10.1109/isgt-europe47291.2020.9248788},
	pages = {749--753},
	booktitle = {2020 {IEEE} {PES} Innovative Smart Grid Technologies Europe ({ISGT}-Europe)},
	publisher = {{IEEE}},
	author = {Ore, Jonathan P. and Groll, Eckhard A.},
	urldate = {2025-07-18},
	date = {2020-10-26},
	langid = {english},
}

@article{ore_case_2021,
	title = {The Case for {DC}: Motivation of Modern Topologies, {DC}-Powered Solutions, and Applications within Residential Environments},
	author = {Ore, Jonathan and Meral, Fatih and Ziviani, Davide and Groll, Eckhard},
	date = {2021},
	langid = {english},
}

@inproceedings{fregosi_comparative_2015,
	location = {Atlanta, {GA}, {USA}},
	title = {A comparative study of {DC} and {AC} microgrids in commercial buildings across different climates and operating profiles},
	url = {http://ieeexplore.ieee.org/document/7152031/},
	doi = {10.1109/icdcm.2015.7152031},
	eventtitle = {2015 {IEEE} First International Conference on {DC} Microgrids ({ICDCM})},
	booktitle = {2015 {IEEE} First International Conference on {DC} Microgrids ({ICDCM})},
	publisher = {{IEEE}},
	author = {Fregosi, Daniel and Ravula, Sharmila and Brhlik, Dusan and Saussele, John and Frank, Stephen and Bonnema, Eric and Scheib, Jennifer and Wilson, Eric},
	urldate = {2025-07-18},
	date = {2015-06},
	langid = {english},
}

@article{sriram_development_2024,
	title = {Development and Comparative Analysis of a Power-over-Ethernet ({PoE}) {DC} Lighting System for Residential Buildings},
	author = {Sriram, Lokesh and Farha, Aaron and Hoess, Andreas and Ziviani, Davide and Groll, Eckhard and Pergantis, Elias and Kircher, Kevin},
	date = {2024},
	langid = {english}
}

@article{anderson_pvlib_2023,
	title = {{PV}lib python: 2023 project update},
	volume = {8},
	rights = {http://creativecommons.org/licenses/by/4.0/},
	issn = {2475-9066},
	url = {https://joss.theoj.org/papers/10.21105/joss.05994},
	doi = {10.21105/joss.05994},
	shorttitle = {pvlib python},
	pages = {5994},
	number = {92},
	journaltitle = {{JOSS}},
	author = {Anderson, Kevin S. and Hansen, Clifford W. and Holmgren, William F. and Jensen, Adam R. and Mikofski, Mark A. and Driesse, Anton},
	urldate = {2025-07-18},
	date = {2023-12-22},
	langid = {english},
	note = {Publisher: The Open journaltitle},
}

@article{chacko_dc_2022,
	title = {{DC} nanogrid for Buildings: Study based on experimental investigation of load performance and Annual energy consumption},
	volume = {58},
	rights = {https://www.elsevier.com/tdm/userlicense/1.0/},
	issn = {2214-7853},
	url = {https://linkinghub.elsevier.com/retrieve/pii/S2214785322009002},
	doi = {10.1016/j.matpr.2022.02.267},
	pages = {352--358},
	journaltitle = {Materials Today: Proceedings},
	author = {Chacko, Rani and Thevarkunnel, Adarsh and Lakaparampil, Z.V. and Thomas, Jaimol},
	urldate = {2025-07-18},
	date = {2022},
	langid = {english},
	note = {Publisher: Elsevier {BV}},
}

@article{dastgeer_comparative_2017,
	title = {A Comparative analysis of system efficiency for {AC} and {DC} residential power distribution paradigms},
	volume = {138},
	rights = {https://www.elsevier.com/tdm/userlicense/1.0/},
	issn = {0378-7788},
	url = {https://linkinghub.elsevier.com/retrieve/pii/S0378778816320461},
	doi = {10.1016/j.enbuild.2016.12.077},
	pages = {648--654},
	journaltitle = {Energy and Buildings},
	author = {Dastgeer, Faizan and Gelani, Hassan Erteza},
	urldate = {2025-07-18},
	date = {2017-03},
	langid = {english},
	note = {Publisher: Elsevier {BV}},
}

@article{dastgeer_analyses_2019,
	title = {Analyses of efficiency/energy-savings of {DC} power distribution systems/microgrids: Past, present and future},
	volume = {104},
	rights = {https://www.elsevier.com/tdm/userlicense/1.0/},
	issn = {0142-0615},
	url = {https://linkinghub.elsevier.com/retrieve/pii/S0142061518305623},
	doi = {10.1016/j.ijepes.2018.06.057},
	pages = {89--100},
	journaltitle = {International journaltitle of Electrical Power \& Energy Systems},
	author = {Dastgeer, Faizan and Gelani, Hassan Erteza and Anees, Hafiz Muhammad and Paracha, Zahir Javed and Kalam, Akhtar},
	urldate = {2025-07-18},
	date = {2019-01},
	langid = {english},
	note = {Publisher: Elsevier {BV}},
}

@article{nzoundja_fapi_control_2025,
	title = {Control Strategy for {DC} Micro-Grids in Heat Pump Applications with Renewable Integration},
	volume = {14},
	rights = {https://creativecommons.org/licenses/by/4.0/},
	issn = {2079-9292},
	url = {https://www.mdpi.com/2079-9292/14/1/150},
	doi = {10.3390/electronics14010150},
	pages = {150},
	number = {1},
	journaltitle = {Electronics},
	author = {Nzoundja Fapi, Claude Bertin and Touré, Mohamed Lamine and Camara, Mamadou-Baïlo and Dakyo, Brayima},
	urldate = {2025-07-18},
	date = {2025-01-02},
	langid = {english},
	note = {Publisher: {MDPI} {AG}},
}

@article{frances_modeling_2018,
	title = {Modeling Electronic Power Converters in Smart {DC} Microgrids—An Overview},
	volume = {9},
	rights = {https://ieeexplore.ieee.org/Xplorehelp/downloads/license-information/{IEEE}.html},
	issn = {1949-3053, 1949-3061},
	url = {https://ieeexplore.ieee.org/document/7932955/},
	doi = {10.1109/tsg.2017.2707345},
	pages = {6274--6287},
	number = {6},
	journaltitle = {{IEEE} Trans. Smart Grid},
	author = {Frances, Airan and Asensi, Rafael and Garcia, Oscar and Prieto, Roberto and Uceda, Javier},
	urldate = {2025-07-18},
	date = {2018-11},
	langid = {english},
	note = {Publisher: Institute of Electrical and Electronics Engineers ({IEEE})},
}

@article{vossos-2017-DC-appliances,
    author = {Vossos, Vagelis and Pantano, Stephen and Heard, Ruby and Brown, Rich},
    title = {{DC} Appliances and {DC} Power Distribution: A Bridge to the Future Net Zero Energy Homes},
    year = {2017},
    journaltitle = {9th International Conference on Energy Efficiency in Domestic Appliances and Lighting},
    url = {https://publications.europa.eu/en/publication-detail/-/publication/a270a15c-fb38-11e7-b8f5-01aa75ed71a1/language-en}
}

@article{vossos_adoption_2022,
	title = {Adoption Pathways for {DC} Power Distribution in Buildings},
	volume = {15},
	rights = {https://creativecommons.org/licenses/by/4.0/},
	issn = {1996-1073},
	url = {https://www.mdpi.com/1996-1073/15/3/786},
	doi = {10.3390/en15030786},
	pages = {786},
	number = {3},
	journaltitle = {Energies},
	author = {Vossos, Vagelis and Gerber, Daniel L. and Gaillet-Tournier, Melanie and Nordman, Bruce and Brown, Richard and Bernal Heredia, Willy and Ghatpande, Omkar and Saha, Avijit and Arnold, Gabe and Frank, Stephen M.},
	urldate = {2025-07-21},
	date = {2022-01-21},
	langid = {english},
	note = {Publisher: {MDPI} {AG}},
}

@inproceedings{rodriguez2015overview,
  title={An overview of low voltage {DC} distribution systems for residential applications},
  author={Rodriguez-Diaz, Enrique and Savaghebi, Mehdi and Vasquez, Juan C and Guerrero, Josep M},
  booktitle={2015 IEEE 5th International Conference on Consumer Electronics-Berlin (ICCE-Berlin)},
  pages={318--322},
  year={2015},
  organization={IEEE}
}

@article{watson2020overview,
  title={An overview of {HVDC} technology},
  author={Watson, Neville R and Watson, Jeremy D},
  journaltitle={Energies},
  volume={13},
  number={17},
  pages={4342},
  year={2020},
  publisher={MDPI}
}

@article{wang2022large,
  title={Large-scale renewable energy transmission by {HVDC}: Challenges and proposals},
  author={Wang, Weisheng and Li, Guanghui and Guo, Jianbo},
  journaltitle={Engineering},
  volume={19},
  pages={252--267},
  year={2022},
  publisher={Elsevier}
}

@article{fotopoulou2021state,
  title={State of the art of low and medium voltage direct current ({DC}) microgrids},
  author={Fotopoulou, Maria and Rakopoulos, Dimitrios and Trigkas, Dimitrios and Stergiopoulos, Fotis and Blanas, Orestis and Voutetakis, Spyros},
  journaltitle={Energies},
  volume={14},
  number={18},
  pages={5595},
  year={2021},
  publisher={MDPI}
}

@inproceedings{lorusso2025introduction,
  title={Introduction to Fault-Managed Power},
  author={Lorusso, Dave and Mlyniec, Stanley and Casey, Jonathan},
  booktitle={2025 IEEE International Symposium on Product Compliance Engineering (ISPCE)},
  pages={1--6},
  year={2025},
  organization={IEEE}
}

@article{pergantis2024field,
  title={Field demonstration of predictive heating control for an all-electric house in a cold climate},
  author={Pergantis, Elias N and Al Theeb, Nadah and Dhillon, Parveen and Ore, Jonathan P and Ziviani, Davide and Groll, Eckhard A and Kircher, Kevin J and others},
  journaltitle={Applied Energy},
  volume={360},
  pages={122820},
  year={2024},
  publisher={Elsevier}
}

@article{pergantis2025protecting,
  title={Protecting residential electrical panels and service through model predictive control: A field study},
  author={Pergantis, Elias N and Premer, Levi D Reyes and Lee, Alex H and Liu, Haotian and Groll, Eckhard A and Ziviani, Davide and Kircher, Kevin J and others},
  journaltitle={Applied Energy},
  volume={386},
  pages={125528},
  year={2025},
  publisher={Elsevier}
}

@article{pergantis2024humidity,
  title={Humidity-aware model predictive control for residential air conditioning: A field study},
  author={Pergantis, Elias N and Dhillon, Parveen and Premer, Levi D Reyes and Lee, Alex H and Ziviani, Davide and Kircher, Kevin J},
  journaltitle={Building and Environment},
  volume={266},
  pages={112093},
  year={2024},
  publisher={Elsevier}
}

@article{pergantis2023sensors,
  title={Sensors, storage, and algorithms for practical optimal controls in residential buildings},
  author={Pergantis, Elias N and Sangamnerkar, Anokhi S and Priyadarshan, Ore J and Dhillon, Parveen and Ziviani, Davide and Groll, Eckhard A and Kircher, Kevin J},
  journaltitle={ASHRAE Trans},
  volume={129},
  pages={437--444},
  year={2023}
}

@misc{duke-price,
    author = {{{D}uke {E}nergy {I}ndiana}},
    title = {Residential rates},
    url = {https://www.duke-energy.com/},
    year = {2025}
}

@article{ashrae-climates,
    author = {{American {S}ociety of {H}eating, {R}efrigeration, and {A}ir-{C}onditioning {E}ngineers}},
    title = {Climatic Data for Building Design Standards},
    journal = {{ASHRAE} Standing Project Committee 169},
    year = {2020}
}

@misc{EMerge-Alliance,
    key = {{E}{M}erge {A}lliance},
    title = {About},
    url = {https://www.emergealliance.org/},
    year = {2026}
}

@Article{alshammari2021dcIreland,
AUTHOR = {Alshammari, Meshari and Duffy, Maeve},
TITLE = {Feasibility Analysis of a DC Distribution System for a 6 kW Photovoltaic Installation in Ireland},
JOURNAL = {Energies},
VOLUME = {14},
YEAR = {2021},
NUMBER = {19},
ARTICLE-NUMBER = {6265},
URL = {https://www.mdpi.com/1996-1073/14/19/6265},
ISSN = {1996-1073},
DOI = {10.3390/en14196265}
}

\appendix

\section{Supporting materials}

\begin{figure}
  \centering
  \includegraphics[width=0.95\columnwidth]{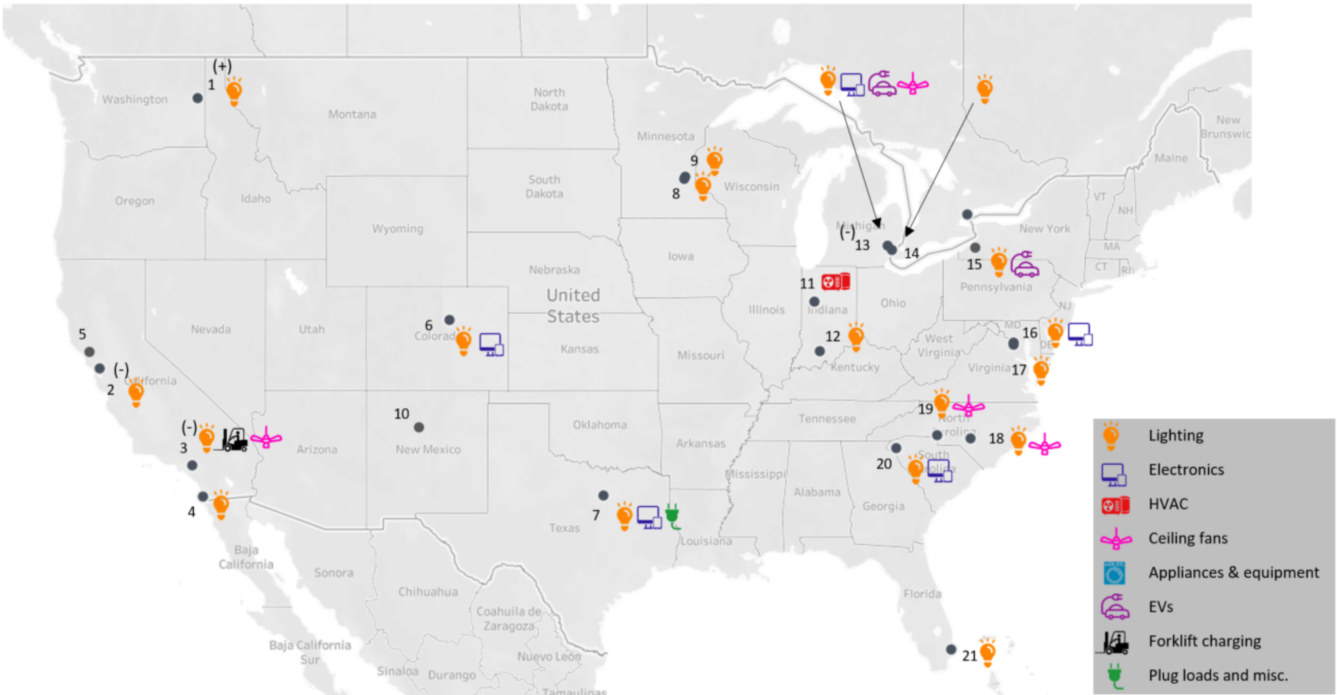}
  \caption{DC distribution deployments in the United States and Canada from 2022 \cite{vossos_adoption_2022}. Reproduced with permission.}
  \label{fig:DC-US-2022}
\end{figure}

Figure \ref{fig:DC-US-2022} shows a map of prior field work on DC-powered devices in the United States and Canada. Only one prior study evaluated DC-powered heat pumps. Table \ref{ta:AC-DC-results-210-240} shows the laboratory results for heat pump capacity, power, and efficiency on AC and DC.

\begin{table}
  \caption{Laboratory test results for AHRI Standard 210/240 (Format: AC, DC)}
  \vspace*{6pt}
  \centering
    \begin{tabular}{@{}l | l l l l l @{}}
        &   \textbf{Thermal} & \textbf{Indoor} & \textbf{Outdoor} & \textbf{Total} & \\ 
    \textbf{Test} & \textbf{capacity} & \textbf{power} & \textbf{power} & \textbf{power} & \textbf{\(\text{COP}\)} \\
     & \textbf{[kW]} &  \textbf{[kW]} & \textbf{[kW]} &  \textbf{[kW]} &  {[-]} \\
    \hline
    A2 & 12.75, 12.44 & 0.372, 0.272 & 3.474, 3.293 & 3.746, 3.565 & 3.40, 3.49 \\
    B2 & 13.78, 13.60 & 0.211, 0.204 & 3.038, 2.840 & 3.155, 3.044 & 4.36, 4.47 \\
    B1 & 4.45, 4.57 & 0.015, 0.028 & 0.684, 0.716 & 0.685, 0.744 & 6.49, 6.14 \\
    Ev & 7.75, 7.48 & 0.092, 0.078 & 1.416, 1.386 & 1.471, 1.464 & 5.27, 5.11 \\
    F  & 5.15, 5.03 & 0.023, 0.033 & 0.529, 0.517 & 0.539, 0.550 & 9.55, 9.15 \\
    H01  & 4.30, 4.58 & 0.072, 0.084 & 0.660, 0.665 & 0.705, 0.749 & 6.09, 6.11 \\
    H11  & 3.51, 3.56 & 0.032, 0.040 & 0.702, 0.714 & 0.716, 0.754 & 4.90, 4.72 \\
    H1n  & 12.86, 12.78 & 0.256, 0.383 & 3.543, 3.374 & 3.799, 3.757 & 3.38, 3.40 \\
    H2v  & 6.98, 6.98 & 0.217, 0.214 & 1.736, 1.727 & 1.954, 1.941 & 3.57, 3.60 \\
    H32  & 10.32, 10.49 & 0.572, 0.544 & 4.292, 4.241 & 4.865, 4.785  & 2.12, 2.19 \\
    \end{tabular}
  \label{ta:AC-DC-results-210-240}
\end{table}

\end{document}